\documentclass{JHEP3}
\usepackage{amsmath}
\usepackage{cite}
\usepackage{epsfig}
\usepackage{graphicx}

\def\be{\begin{equation}}
\def\ee{\end{equation}}
\def\baray{\begin{eqnarray}}
\def\earay{\end{eqnarray}}
\def\ba{\begin{eqnarray}}
\def\ea{\end{eqnarray}}

\def\fid{\dot{\phi}}

\title{Rapid Tunneling and Percolation in the Landscape}

\author{Saswat Sarangi $^1$, Gary Shiu $^2$, Benjamin Shlaer $^3$\\
$^1$ Institute of Strings, Cosmology and Astroparticle Physics,
Department of Physics, Columbia University, New York, NY 10027, USA\\
{\tt sash@phys.columbia.edu} \\
$^2$  Department of Physics, University of Wisconsin, Madison, WI 53706, USA \\
{\tt shiu@physics.wisc.edu} \\
$^3$ Physics Department, University of Colorado, Boulder, CO 80309, USA \\
{\tt bshlaer@pizero.colorado.edu}}
\date{\today}

\abstract{Motivated by the possibility of a string landscape, we reexamine tunneling 
of a scalar field across single/multiple barriers. Recent investigations
have suggested modifications to the usual picture of false vacuum decay
that lead to efficient and rapid tunneling in the landscape when certain conditions are met. 
This can be due to stringy effects (e.g. tunneling via the DBI action), or 
by effects arising due to the presence of multiple vacua (e.g. resonance tunneling).
In this paper we discuss both DBI tunneling and resonance tunneling.
We provide a QFT treatment of resonance tunneling using the Schr\"odinger functional approach.
We also show how DBI tunneling for supercritical barriers can naturally lead to conditions
suitable for resonance tunneling. We argue using basic 
ideas from percolation theory that tunneling can be rapid 
in a landscape where a typical vacuum has multiple decay channels  
and discuss various cosmological implications. This rapidity vacuum decay
can happen even if there are no resonance/DBI tunneling enhancements, solely
due to the presence of a large number of decay channels.
Finally, we consider various ways 
of circumventing a  recent no-go theorem for resonance tunneling in quantum field theory. 
}

                                         
\preprint{MAD-TH-07-09}

\begin{document}

\section{Introduction}
Various theoretical considerations have led to a picture of string
theory with a gigantic number of solutions, referred to as the
Landscape of vacua \cite{Susskind:2003kw}. This picture was
motivated in part by the construction of cosmological vacua in
\cite{Kachru:2003aw, Kachru:2003sx}, and the dense
``discretuum'' in \cite{Bousso:2000xa}. Although a detailed
knowledge of the structure of the landscape is lacking, it is usually
thought of as a complicated potential energy function in a multidimensional moduli space with a multitude
of local minima separated by barriers of various heights. One of these minima corresponds to the vacuum where
the universe exists today and confers it with its various characteristics (the cosmological
constant, broken supersymmetry, the standard model etc).  Simple models of the landscape such as those discussed in \cite{Bousso:2000xa,Arkani-Hamed:2005yv}
have different values of the cosmological constant and dimensionful couplings (such as Higgs mass) in the different
vacua. The existence of a large number of vacua may
shed some light on the cosmological constant problem \cite{Bousso:2000xa}. 

 Given an initial de Sitter vacuum with a string scale (or Planck scale)
cosmological constant, successive tunneling events to vacua with lower vacuum energies relax the cosmological constant. While this
relaxation mechanism is happening, the universe at large scales eternally inflates and the global structure becomes very complicated
\cite{Linde:2005ht}, with various regions of spacetime inflating in various vacua. These regions are separated by domain walls. Eventually, some region of spacetime tunnels down to a metastable vacuum with a low cosmological constant, long 
lifetime and standard model fields.  On the largest scales (far beyond scales which could ever be observed) the universe is not
in our vacuum, and (assuming metastability) our vacuum will eventually catastrophically recede outside of our horizon, leaving a new and certainly hostile vacuum for us to observe.   
Tunneling is an important ingredient in this picture of the early universe.
The universe samples the different sites of the landscape
by some form of tunneling (two common well studied forms of tunneling being Coleman De Luccia and Hawking Moss instantons).  

Recent attempts at finding a cosmological measure for the landscape \cite{Vilenkin:2006xv}
employ tunneling as the mechanism that allows the universe to sample various vacua. In fact, the occurrence
of eternal inflation in the landscape itself depends on the mechanism of tunneling.\footnote{ 
It is not clear if slow roll eternal inflation is generic in the string 
inspired inflationary scenarios.  In some scenarios it is absent\cite{Chen:2006hs, Arkani-Hamed:2007ky}.}
Tunneling mechanisms that have been widely 
studied in field theories \cite{Coleman:1977py,Coleman:1980aw,Hawking:1981fz} display an
exponential suppression of the tunneling rate with an increase in the height of the barrier
between the vacua. The expectation in the string landscape has been that the tunneling rates
will be far suppressed compared to the rate of expansion of the universe during the de Sitter
phase due to the generically string scale barriers. As a result, most regions of the universe eternally
inflate, i.e. most regions are stuck in their various vacua with various positive vacuum energies.
Certain regions of spacetime occasionally tunnel to regions of the landscape with a flat potential
conducive to slow roll inflation. Other regions that tunnel to vacua
with negative cosmological constant are unstable to gravitational collapse (big crunch singularities) \cite{Coleman:1980aw}.
  
Because tunneling plays such a central role in the string landscape, it is important to fully understand 
it in such a setting.  We assume from the start that transitions in the landscape have a four-dimensional effective field theory 
description\footnote{We thank Shanta de Alwis for emphasizing this point to us \cite{Alwis:2006cb}.}.  In this paper we reexamine and expand the scope of tunneling in the context of the 
string landscape. In particular, we discuss two modes of tunneling that differ from the
traditional view of tunneling in that they can lead to greatly enhanced transition rates even for tall barriers.
These two modes are resonance tunneling \cite{Tye:2006tg} and DBI tunneling \cite{Brown:2007ce}.
A crucial feature of these two modes of tunneling is that both of them make distant vacua accessible for tunneling.
That is to say, in addition to the usual tunneling between neighboring vacua (those vacua which share a common barrier), tunneling can now occur
at a relatively unsuppressed rate between vacua that are separated by other vacua (i.e. vacua which are not immediate neighbors).
We use the term {\it Efficient Tunneling} for this enhancement of tunneling rate that can make transitions to far away
vacua relatively unsuppressed \cite{Henry}. This efficiency in tunneling (access to far away vacua) increases the number 
of decay channels from a given vacuum. When the number of decay channels
becomes large enough, tunneling can become rapid (i.e. transitions from one vacuum to another can occur with order one 
tunneling rate). Efficient tunneling is a necessary (but not sufficient) condition for rapid tunneling.
As we shall discuss in this paper, the string landscape can naturally provide an arena where efficient tunneling can also
become rapid.

This paper consists of the following parts. In Sec. \ref{resonance} we discuss resonance tunneling using
semiclassical methods. In the semiclassical approximation one can describe resonance tunneling in QFT models of
the landscape. To this end, we use the methods developed by Ref \cite{Banks:1973ps,Gervais:1977nv,Bitar:1978vx} to study quantum 
tunneling in QFT. We discuss the importance of {\it coherence} for resonance tunneling to take place in
a certain part of the landscape. Sec. \ref{dbi} of this paper involves
a discussion of DBI tunneling based on the analysis done in \cite{Brown:2007ce}. This provides a description
of tunneling phenomena in the open string landscape where the dynamics is described by the DBI action. We shall
see that both resonance tunneling and DBI tunneling involve major modifications to the Coleman-De Luccia result and
lead to an enhancement of tunneling rates. Further, we shall see an interesting connection between resonance tunneling
and DBI tunneling: DBI tunneling for supercritical barriers naturally leads to a condition 
necessary for resonance tunneling.  In Sec. \ref{percolate} of this paper we 
use results from percolation theory to show how these efficient tunneling mechanisms, even if 
rare occurrences in the landscape, can still make a significant qualitative change to the picture. It is expected
that a typical vacuum in the landscape will have a large number of decay channels. In flux compactifications, for
example, 
a large number of decay channels is a result of many independent fluxes (cycles).
The possibility of efficient tunneling, 
even if rare, together with the existence of a large number of flux directions can have 
nontrivial implications. In particular, in the presence of a large number of decay channels efficient tunneling
can lead to rapid tunneling. Even though resonance/DBI tunneling offer novel ways of vacuum decay,
they are not entirely necessary for rapid tunneling. In Sec. \ref{cheby} we show how the presence
of a large number of decay channels, as expected in the flux landscape, can make even the usual
vacuum decay a rapid process.

In Sec. \ref{discuss} we discuss the various cosmological implications of our investigations.
Here we summarize them: 
\begin{itemize}
\item The standard  Coleman - De Luccia tunneling rate is responsible for the conventional view of 
the string landscape as an eternally inflating ensemble of meta-stable vacua.  In this framework, 
tunneling to nearby vacua is exponentially suppressed, and transitions to vacua with larger cosmological constant, or to vacua more distant in field space incur additional exponential suppression.  Thus each site in the landscape has some degree of physical viability.  While disconcerting for most purposes, the enormous size of the accessible landscape can be appealing for reasoning based on the 
anthropic principle or similar statistical methods.  This picture may undergo striking modification when efficient tunneling and the large dimensionality of the landscape are fully taken into account.  In this new picture, two effects will alter any statistical framework.  Firstly, certain decay channels are found to be vastly preferred over every other channel (sometimes over even those much closer in field space).  Secondly, many vacua formerly thought to be meta-stable can become essentially unstable due to rapid tunneling.  These two features can effectively remove large portions of the landscape, and may limit the domain of eternal inflation.

\item Fast tunneling needs to be shut off for vacuum sites with low cosmological constant (comparable to today's value).  This could be achieved in two ways:  by removing the possibility of efficient tunneling, or by reducing the number of nearest neighbors (dimensionality) in the field space.   If this shutting-off of rapid tunneling is sudden, then this can
be a natural explanation for the cosmological constant problem. 
The universe begins at some vacuum site with a large (string scale or Planck scale)
vacuum energy. However, rapid tunneling makes the universe quickly decay down to a (non-eternal) inflationary vacuum site 
that ends with reheating and a low vacuum energy.  In \cite{Henry}, RG group arguments are used to give an explanation 
for the necessary sudden shut-off of rapid tunneling.  It is known that gravitational effects on vacuum decay can elegantly prevent\cite{Coleman:1980aw} certain transitions to negative cosmological constant vacua, but they are of no help as a shut-off mechanism for the small domain wall tensions which coincide with rapid tunneling.  Instead, one must assume another mechanism is able to induce the shut off of rapid tunneling which our current vacuum enjoys.

\item Fast tunneling gives a natural solution to the {\it empty universe} problem inherent in Brown-Tietelboim and
Bousso-Polchinski scenarios. The empty universe problem arises due to the exponentially large lifetime of
vacua from which the universe decays down to today's vacuum. The exponentially large time required for the decay dilutes
away all the radiation/matter content of the universe. Bousso-Polchinski used the Hubble fluctuations to perch a classical inflaton 
field for subsequent slow roll inflation as a plausible solution to this problem.  However, rapid tunneling provides a 
similar natural solution due to the short lifetime of vacua with high vacuum energies.  This is because rapid tunneling can resemble
classical inflation. 
\item In some scenarios, rapid tunneling can make parts of the landscape accessible during the last $60$ e-folds of inflation. 
This can lead to the formation of domain walls, the phenomenology of which 
has been
discussed in \cite{Davoudiasl:2006ax}. 
\end{itemize}

\section{Resonance Tunneling} \label{resonance}

There are several examples where the quantum mechanical behavior of scalar fields plays an essential role during periods
of cosmological inflation.  Arguably the most familiar and important example are the inflaton's quantum fluctuations which
lead to the density perturbations in our universe.  Guth and Pi did
a proper quantum mechanical treatment of the scalar field in \cite{Guth:1985ya}. The quantum treatment of a scalar
field in the stringy landscape is further in that spirit.

In many situations, a quantum field theoretic (QFT) 
problem reduces to a simple quantum mechanical (QM) problem. For example, to understand aspects of the
behavior of the inflaton field, it suffices under many circumstances to look at the analogous quantum mechanical problem \cite{Guth:1985ya}. The main
complication that arises within QFT is the presence of an infinite number of Fourier modes. While the dynamics of QM
is contained within the Schr\"odinger equation for the single particle wave function, the QFT problem is described by 
the Schr\"odinger functional equation, and the field is described by a wavefunctional. For a free field theory the wavefunctional
factorizes to an infinite number of wavefunctions - one for each Fourier mode, and the QFT is then neatly described by an infinite
number of Schr\"odinger equations.
However, when the field theory is an interacting one (as it must 
be for tunneling to exist), these infinitely numerous 
of Schr\"odinger equations become coupled with each other. Describing
tunneling in a QFT might therefore appear to be a very complicated problem. Banks, Bender and Wu \cite{Banks:1973ps} extended
the simple WKB tunneling picture in QM involving one degree of freedom to multiple degrees of freedom. Others \cite{Gervais:1977nv,
Bitar:1978vx} extended the analysis to an infinite dimensional configuration space (QFT). The main idea underlying this extension
from QM to QFT is  
that within the purview of the semiclassical approximation, tunneling under a barrier occurs in a tubular region
within the infinite dimensional configuration space which  
can be described as a solution to a Euclidean classical equation of motion.
This trajectory can be elegantly described using Euclidean path integrals, as in \cite{Coleman:1977py}. This enormous simplification
of the problem under the semiclassical scheme makes the problem of tunneling tractable in QFT.  In the following subsections we shall use the Schr\"odinger formalism to
apply this basic idea to resonance tunneling within QFT. 

\subsection{Resonance tunneling in quantum mechanics} \label{QM}

We first summarize resonance tunneling in quantum mechanics. Consider a particle moving in one dimension $x$ under the influence
of a potential $V(x)$. Fig.($1$) shows a potential with minima A, B and C separated by barriers. The wavefunction of the 
particle, $\psi(x)$, can be obtained by solving the Schr\"odinger equation $H \psi = E \psi$. In the classically allowed region, 
via the WKB approximation, this gives
\baray
\label{ca}
\psi_{L,R}(x) \simeq \frac{1}{\sqrt{k(x)}}\exp \left( \pm  i \int^x dx k(x) \right)\,\, ,
\earay
where $k(x) = \sqrt{\frac{2m}{\hbar ^2}(E - V(x))}$. The $\pm$ signs correspond to the left moving and the right moving solutions.
A general solution in the classically allowed region will be a superposition $\psi(x) = \alpha_L \psi_L(x) + \alpha_R \psi_R(x)$.
Similarly, in the classically forbidden region the solutions are
\baray
\label{cf}
\psi_{\pm}(x) \simeq \frac{1}{\sqrt{\kappa(x)}}\exp \left( \pm  \int^x dx \kappa(x) \right)\,\, ,
\earay
where $\kappa(x) = \sqrt{\frac{2m}{\hbar ^2}(V(x)-E)}$, and the $\pm$ signs correspond to the growing and the decaying solutions.
A general solution in the forbidden region will again be a linear superposition $\psi(x) = \alpha_+ \psi_+(x) + \alpha_- \psi_-(x)$.
When a situation involves the propagation of a particle in both the classically allowed and classically forbidden regions,
the full solution has the form Eq.(\ref{ca}) in the allowed region and Eq.(\ref{cf}) in the forbidden region. Since these both
represent the same solution in different regions, the coefficients $\alpha_{L}$, $\alpha_{R}$, $\alpha_+$ and $\alpha_-$ are 
related by the {\it connection formulas}.
For concreteness, let us consider the problem of the tunneling of the particle from vacuum A to vacuum B in Fig.($1$). Let the coefficients
of the components $\psi_{L,R}$ in vacuum A be $\alpha_{L,R}$ and the coefficients in vacuum B be $\beta_{L,R}$. These coefficients are related
to each other by the following matrix expression which can be derived from the connection formulae (see, e.g., \cite{Merzbacher}).
\baray
\label{match}
\left(
\begin{array}{c}
\alpha_R \\
\alpha_L \end{array}
\right) 
=
\frac{1}{2}\left(
\begin{array}{cc}
\Theta + \Theta^{-1}    & i(\Theta - \Theta^{-1}) \\
-i(\Theta - \Theta^{-1})& \Theta + \Theta^{-1} \end{array}
\right)
\left(
\begin{array}{c}
\beta_R \\
\beta_L \end{array}
\right) 
\earay 
where $\Theta$ is given by
\baray
\Theta \simeq 2 \exp \left( \frac{1}{\hbar}\int_{x_1}^{x_2} dx \sqrt{2m(V(x) - E)} \right) \,\, ,
\earay
$x_1$ and $x_2$ being the classical turning points. To study tunneling of a wave packet from vacuum A to vacuum B, we keep 
only a right moving component in vacuum B (i.e. there is no particle incident on the barrier from the right), $\beta_L = 0$.
The tunneling probability is given by
\baray
T_{A \to B} = |\frac{\beta_R}{\alpha_R}|^2 = 4 \left( \Theta + \frac{1}{\Theta} \right)^{-2} \simeq \frac{4}{\Theta ^2} \,\, ,
\earay 
as $\Theta$ is usually an exponentially suppressed quantity.

Now we consider the possibility of resonance tunneling.  We are interested in the tunneling of the particle from vacuum A to C 
via B. The two intervening barriers (between A and B and between B and C) contain the classically forbidden regions while 
B is classically allowed. One can again derive the relation between the coefficients $\alpha_{L,R}$ (in vacuum A) and $\gamma_{L,R}$
(in vacuum C) and express it in the following matrix form.
\baray
\left(
\begin{array}{c}
\alpha_R \\
\alpha_L \end{array}
\right) 
=
\frac{1}{2}\left(
\begin{array}{cc}
\Theta + \Theta^{-1}    & i(\Theta - \Theta^{-1}) \\
-i(\Theta - \Theta^{-1})& \Theta + \Theta^{-1} \end{array}
\right) \times
\left(
\begin{array}{cc}
e^{-\frac{i}{\hbar}W} & o \\
0                     & e^{\frac{i}{\hbar}W} \end{array}
\right) \nonumber \\
\times
\left(
\begin{array}{cc}
\Phi + \Phi^{-1}    & i(\Phi - \Phi^{-1}) \\
-i(\Phi - \Phi^{-1})& \Phi + \Phi^{-1} \end{array}
\right) \times
\left(
\begin{array}{c}
\gamma_R \\
\gamma_L \end{array}
\right) 
\earay
where
\baray
\label{phi}
\Phi \simeq 2 \exp \left( \frac{1}{\hbar} \int_ {x_3}^{x_4} dx \sqrt{2m\left(V(x)-E\right)}  \right)   
\earay
and
\baray
\label{W}
W = \int_{x_2}^{x_3} dx  \sqrt{ 2m(E - V(x))}\,\, , 
\earay
with $x_3$ and $x_4$ the turning points on the barrier between B and C.
The transmission probability from A to C, via B, is now given by
\baray
\label{AtoC}
T_{A \to C} = 4 \left( \left( \Theta \Phi + \frac{1}{\Theta \Phi}\right)^2 \cos ^2 W + \left( \frac{\Theta}{\Phi} + \frac{\Phi}{\Theta} \right)^2  \sin^2 W \right)^{-1}\,\,.
\earay
In general this will be a small quantity. However, when the so called {\it resonance condition} is satisfied, 
$T_{A \to C}$ can be enhanced. This happens when both $W$ satisfies the ``bound state condition"  
\baray
\label{reson}
W = \left( n + \frac{1}{2} \right) \pi, \quad n = 0, 1, 2, 3, ...
\earay
such that $\cos W = 0$, and $\Theta \to \Phi$, i.e. $T_{A \to B} \to T_{B \to C}$.
Under this resonance condition the transmission probability from vacuum A to C is greatly enhanced, $T_{A \to C} \to 1$.
Physically this enhancement is due to the constructive interference that occurs in the classically allowed region in vacuum B
when the resonance condition (Eq.(\ref{reson})) is satisfied. 

One can estimate the probability that the resonance
condition is actually satisfied.  We define $\Delta E$ to be the energy window  
where the resonance condition is
roughly satisfied.  This can be the range of $E$ where $T_{A \to C}$ varies between $1/2$ and $1$ (i.e. it is the width
at half maximum).  Let $E_0$ be the approximate energy difference between a typical pair of 
neighboring resonances. Then the probability for the
occurrence of resonance is $\Delta E / E_0$. Typically this is an exponentially small quantity, of the order of
$T_{A \to B}$ and $T_{B \to C}$. More precisely \cite{Tye:2006tg}, $\Delta E / E_0 \simeq (T_{A \to B} + T_{B \to C})/2\pi$.

\subsection{Resonance tunneling in quantum field theory} \label{QFT}

\begin{figure}
\begin{center}
\includegraphics[width=9cm]{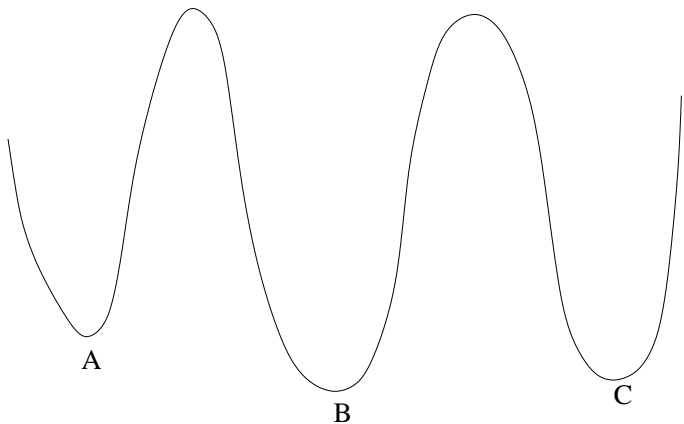}
\vspace{0.1in}
\caption{Vacua A, B and C separated by barriers in the landscape.}
\label{fig:abc}
\end{center}
\end{figure}

Tunneling in QFT has been well studied in the semiclassical regime using both the Euclidean path integral approach \cite{Coleman:1977py}
 and the functional Schr\"odinger equation approach \cite{Gervais:1977nv, Bitar:1978vx}. The tunneling between two vacua is often well described by a WKB calculation.    
 The tunneling rate is generically suppressed as the height or width of the 
 barrier between the two vacua is increased,  and a typical tunneling rate is exponentially small.  The tunneling
 proceeds via bubble nucleation, and if the true vacuum interior is large enough, the bubble wall grows and quickly attains relativistic velocities. 
 The picture might be more complicated when there are more
 features to the shape of the potential. Such considerations are especially important now in view of the string landscape.  As recently
 discussed in \cite{Tye:2006tg, Davoudiasl:2006ax}, the presence of multiple vacua in the string landscape can give rise to
 various interesting effects arising due to efficient tunneling.  One expects a large multiple vacua separated by high barriers to be a
 generic feature of the landscape (see \cite{Danielsson:2006xw}). In \cite{Tye:2006tg}, a simple landscape  is considered consisting of vacua
A, B and C,  with B being the intermediate between vacua A and C in the field space, as shown in Fig.(\ref{fig:abc}).   
When the resonance 
condition is satisfied, one expects the transmission probability to be of order unity to go from vacuum A to vacuum C via vacuum B.  This means
that any other vacua will have negligible branching fractions.   The passage of the
{\it wave packet} through vacuum B is quantum mechanical and leads to a constructive interference that is responsible for
making the transmission probability close to one.  This, of course, can have nontrivial effects for the landscape and early cosmology
(see e.g. \cite{Tye:2006tg,Davoudiasl:2006ax}).

In this section we begin 
a field theoretic analysis of resonance tunneling. The analysis in \cite{Tye:2006tg} is quantum mechanical
and is based on solving the Schr\"odinger equation for a single particle. However, to understand the problem in QFT we must look at the functional
Schr\"odinger equation and analyze the problem using WKB methods.  The semiclassical approximation to the Euclidean path integral cannot capture the effect of resonance tunneling.  
The Euclidean method captures the under-the-barrier effect between two vacua.  However, for resonance 
tunneling, the motion in the intermediate classically allowed region is important.  Although the Euclidean instanton technique 
is quite elegant in studying the problem of barrier penetration and false vacuum decay, the same problem can be studied using WKB methods
in the functional Schr\"odinger equation formalism \cite{Gervais:1977nv,Bitar:1978vx, Vachaspati:1991tq}. These references, 
in turn, build on the ideas developed by Banks, Bender and Wu in \cite{Banks:1973ps} to study quantum tunneling in dynamical systems with a large number of degrees of freedom.  We shall adapt the methods spelt out in \cite{Gervais:1977nv,Bitar:1978vx} to study resonance tunneling in QFT. This method captures both the classically forbidden region as well as the classically allowed region.  It is therefore especially suited to study the phenomenon of resonance tunneling which  involves both forbidden and allowed regions in the field space in an essential way.   
 
The basic idea is based on constructing the WKB wavefunctions for the system in the multidimensional 
configuration space (corresponding to the various Fourier modes). This is simple for a system with a single 
degree of freedom (as in quantum mechanics for a single particle). The work done in \cite{Banks:1973ps,Gervais:1977nv,
Bitar:1978vx} generalizes the WKB construction to systems with many degrees of freedom by arguing that under 
semiclassical approximation, barrier penetration happens mostly in small tubes within the configuration space around 
certain classical solutions, so that the WKB approximation is still essentially one dimensional. These classical 
paths correspond to classical solutions with Wick rotated (imaginary) time. The order $\hbar$ corrections can be calculated by 
considering small fluctuations about this classical trajectory. Ref \cite{Gervais:1977nv} calculates
WKB wavefunctions to the first two orders in $\hbar$ systematically. 

One begins with the full Hamiltonian $H(\phi(x), \dot{\phi(x)}) = H(\phi(x), \frac{\hbar}{i}\delta / \delta \phi(x))$, and writes down the
Schr\"odinger functional equation for the wave functional $\Psi(\phi(x))$ of the field.  Using the WKB approximation one may solve
for this wave functional.  The lowest order WKB approximation gives the equation of motion that determines the most probable escape path (MPEP) along which
tunneling proceeds. The next order WKB gives corrections due to gaussian fluctuations around this path.   
In the remainder of this section we construct the WKB solutions in both the classically allowed and classically forbidden regions. 

\subsection{Configuration space}

The configuration space of the field $\phi(x)$ is of infinite dimension.  (In our $3+1$ dimensions, $x$ corresponds to the {\em spatial}  
directions.)  We shall use $\lambda$ to parameterize a one-dimensional path in this configuration space.  According to \cite{Banks:1973ps},
in the semiclassical approximation, tunneling happens along a ``most probable escape path" (MPEP) that can be parameterized by the variable $\lambda$. This makes the problem tractable. The nearby paths will contribute quantum corrections that can be systematically
calculated. The parameter $\lambda$ has nothing to do with the imaginary time that one uses in Euclidean techniques, it is just a parameter
that describes the MPEP. 

A point in this configuration space is labeled by a parameter $\lambda$. The
coordinates of this point are given by the complete function  
$\{\phi(x;\lambda)\} \equiv \{\phi(x_1;
\lambda), \phi(x_2; \lambda), ...\}$. Here  $\{ x_1, x_2, ...\}$ span all
of space and, hence, there is an continuous infinity of these. 
A trajectory is given by, for example,
$0 < \lambda < 1$, and as $\lambda$ continuously varies, so does the field 
configuration given by $\phi(x, \lambda)$. A line element $dr$ along the path 
$\phi(x, \lambda)$ is given by
\baray
\label{path}
(dr)^2 = \int d^3x (d \phi(x))^2 = (d\lambda)^2 \int d^3x \left( \frac{\partial \phi(x)}{\partial\lambda} \right)^2\,\,.
\earay
So if  the length along a path, $r$, is used to parameterize the path,
then $\lambda = r$, and $\int d^3x \left(\partial \phi/ \partial r  \right)^2
=1$. One can define the tangent vector along the path length $dr$ and 
a vector orthogonal to the path $\phi(x,r)$ as
\baray
\delta \phi_{tan} &=& dr \frac{\partial \phi}{\partial r} \nonumber \,\,,\\
\delta \phi_{\perp} &=& \delta \phi - a dr \frac{\partial \phi}{\partial r}\,\,,
\earay
where $\delta \phi$ is an arbitrary variation, and $a$ is chosen such that
\baray
\int d^3 x \delta \phi_{\perp} \left( \frac{\partial \phi}{\partial r} \right)
= 0\,\, ,
\earay
which implies
\baray
a dr = \int d^3x \delta \phi \left( \frac{\partial \phi}{\partial r} \right)\,\,.
\earay
In general, there are an infinity of such orthogonal vectors at
every point $\phi(x,r)$ of the trajectory.  For every such direction
$n(y)$ we may define a unit vector as
\baray
\hat{\phi}_{n(y)}(x,r) = \frac{\delta \phi_{\perp}}{[\int d^3x(\delta 
\phi_{\perp})^2 ]^{1/2}}\,\,,
\earay
where $\hat{\phi}_{n(y)}(x,r)$ depends on $r$, and the variable 
$y$ denotes the coordinates of the direction $n(y)$. The orthogonal
variation can now be formally written as
\baray
\label{perp}
\Delta \phi_{\perp} = \int d^3 y n(y) \hat{\phi}_{n(y)}(x,r)\,\,.
\earay
Therefore, for a point $\phi(x)$ near the path defined by $\phi(x,r)$,
there exists a value of $r$ such that
\baray
\phi(x) = \phi(x,r) + \Delta \phi_{\perp}(x,r)\,\,,
\earay
where $\Delta \phi_{\perp}(x,r)$ is given by Eq.(\ref{perp}).

\subsection{Functional Schr\"odinger equation}
The Hamiltonian for the scalar field $\phi(x)$ is given by
\baray
H = \int d^3x \left( \frac{1}{2}\dot{\phi}^2 + \frac{1}{2}(\nabla \phi)^2 + V(\phi)  \right)\,\,,
\earay 
where we assume that $V(\phi)$ has the shape shown in Fig.(\ref{fig:abc}). To quantize the field theory, one notes that the commutator of $\dot{\phi}$ and $\phi$ should be given by $[\dot{\phi}(x), \phi(x')]= i\hbar \delta^3 (x-x')$, and so 
 $\dot \phi(x)$ is replaced by the operator $- i\hbar \delta/\delta \phi(x)$.
The time independent functional Schr\"odinger equation is given by
\baray
H \Psi(\phi) = E \Psi(\phi) \,\,,
\earay
where the Hamiltonian operator is
\baray
H = \int d^3x \left( - \frac{\hbar^2}{2} \left( \frac{\delta}{\delta \phi(x)} \right)^2+ \frac{1}{2}(\nabla \phi)^2 + V(\phi)  \right)\,\,.
\earay
Here $\Psi(\phi(x))$ has the usual probabilistic interpretation, giving a measure of the probability of the occurrence of a field configuration
$\phi(x)$.

\subsection{WKB approximation}
We assume a solution of the form $\Psi(\phi)= A e^{\frac{i}{\hbar}S(\phi)}$, where $A$ is a constant. The Schr\"odinger equation then looks like
\baray
\label{hj}
\int d^3x \left(  -\frac{\hbar^2}{2} \left[ \frac{i}{\hbar}\frac{\delta^2 S(\phi)}{\delta \phi^2} - \frac{1}{\hbar^2}\left( \frac{\delta S(\phi)}{\delta \phi}\right)^2 + \frac{1}{2}(\nabla \phi)^2 + V(\phi)\right]   \right) e^{\frac{i}{\hbar}S(\phi)} = E e^{\frac{i}{\hbar}S(\phi)}\,\,.
\earay
Expanding $S(\phi)$ in powers of $\hbar$, $S(\phi) = S_{(0)}(\phi) + \hbar S_{(1)}(\phi)+ ...$, and comparing
terms with similar powers of $\hbar$ in the above equation leads to the following equations 
\baray
\label{hj1}
\int d^3x \left[ \frac{1}{2}\left(\frac{\delta S_{(0)}(\phi)}{\delta \phi}\right)^2 + \frac{1}{2}(\nabla \phi)^2 + V(\phi) \right] = E, 
\earay
\baray
\label{hj2}
\int d^3x \left[ -i\frac{\delta^2 S_{(0)}(\phi)}{\delta \phi^2} + 2 \frac{\delta S_{(0)}(\phi)}{\delta \phi}\frac{\delta S_{(1)}(\phi)}{\delta \phi} \right] = 0,
\earay
\baray
etc. \nonumber   
\earay
 Eqns.(\ref{hj1}, \ref{hj2}) represent an infinite set of coupled nonlinear equations in the infinite dimensional configuration space of the
 field $\phi(x)$. To first approximation, however, it is possible to reduce these equations to a one dimensional equation.  To this end, first define the potential $U(\phi)$ related to $V(\phi)$ as follows
\be 
U(\phi) = \int d^3x \left( \frac{1}{2}(\nabla \phi)^2 + V(\phi) \right).
\ee 
 The classically forbidden region is defined as the region where $U(\phi) > E$, and the classically allowed region is where $U(\phi) < E$.  
  Let us first look at the WKB solutions for the classically forbidden region (region II of Fig.(\ref{fig:U123})). The analysis for the classically 
  allowed region will be similar. The crux of the argument in \cite{Banks:1973ps} is that there is a trajectory in the configuration 
  space of the field - the most probable
 escape path - perpendicular to which any variations of $S_{(0)}$ vanish, and along which the variations of $S_{(0)}$ are nonvanishing. If we use
 the parameter $\lambda$ to describe this one dimensional path in the infinite dimensional configuration space, then the most probable escape
 path is given by $\phi_0(x, \lambda)$, and $\delta S_{(0)}/\delta \phi$ is nonvanishing only along this path. If such a path exists, then we
 can convert the hopelessly complicated set of Eqns(\ref{hj1}, \ref{hj2}...) to a very simple one dimensional equation. This can be done as
 follows. The most probable escape path satisfies
 \baray
 \label{mpep}
 \frac{\delta S_{(0)}}{\delta \phi_{||}} \mid_{\phi_0(x,\lambda)} &=& C(\lambda) \frac{\partial \phi_0}{\partial \lambda}, \nonumber \\
 \frac{\delta S_{(0)}}{\delta \phi_{\perp}} \mid_{\phi_0(x,\lambda)} &=& 0\,\,.
 \earay
Further, $C(\lambda)$ is related to $\partial S_{(0)}/\partial \lambda$, and can be eliminated.  Note that along the most probable escape path one can write
\baray
\label{mpep1}
\frac{\partial S_{(0)}}{\partial \lambda} = \int d^3x \frac{\delta S_{(0)}}{\delta \phi_{||}}\mid_{\phi_0(x,\lambda)} \frac{\partial \phi_0}{\partial \lambda}\,\,,
\earay  
so
\baray
\label{C}
C(\lambda) = \frac{\partial S_{(0)}}{\partial \lambda} \left( \int d^3x \left[\frac{\partial \phi_o}{\partial \lambda}\right]^2\right)^{-1}\,\,.
\earay 
Combining Eq(\ref{C}) and Eq(\ref{mpep}) with Eq(\ref{hj1}), one gets the WKB equation
\baray
-\frac{1}{2}\left( \int d^3x \left[\frac{\partial \phi_o}{\partial \lambda}\right]^2\right)^{-1} \left( \frac{\partial S_{(0)}}{\partial \lambda} \right)^2
= U(\phi_0(x,\lambda)) - E\,\,.
\earay
Finally, using the definition of the path length $dr$ in the field configuration space as given by Eq(\ref{path}), one reduces the field theory
WKB approximation to the one dimensional  
\baray
\label{hj1*}
-\frac{1}{2}\left( \frac{\partial S_{(0)}}{\partial r} \right)^2 = U(r) - E,
\earay
where $U(r)$ is the potential along the most probable escape path given by
\baray
U(r)= \int d^3x \left( \frac{1}{2}(\nabla \phi_0(x,r))^2 + V(\phi_0(x,r)) \right)\,\,.
\earay
Similarly, Eq.(\ref{hj2}) for $S_{(1)}$ becomes the one dimensional equation
\baray
\label{hj2*}
i\frac{\partial^2{S_{(0)}}}{\partial r^2} - 2 \frac{\partial S_{(0)}}{\partial r} \frac{\partial S_{(1)}}{\partial r} = 0.
\earay
The familiar solutions to Eq(\ref{hj1*}) and Eq.(\ref{hj2*}) are
\baray
\label{soln}
S_{(0)} &=& i \int_0^r dr' \sqrt{2[U(r')-E]}, \nonumber \\
S_{(1)} &=& \frac{i}{2}\ln \sqrt{2[U(r)-E]}. 
\earay
The WKB wavefunctional effectively reduces to a WKB wavefunction and can now be written as
\baray
\label{wkb}
\Psi_{II}(\phi_0(x,r)) = \frac{A}{{(2[U(r)-E])}^{1/4}} \exp \left( -\frac{1}{\hbar}\int_0^r dr' \sqrt{2[U(r')-E]} + ...\right)\,\,,
\earay
where the II subscript corresponds to the classically forbidden region II in Fig.(\ref{fig:U123}).
One can also determine the equation for the most probable escape path.  This path corresponds to the 
Hamilton-Jacobi variational equation
\baray
\delta S_{(0)} = \delta \int_0^r dr' \sqrt{2[U(r')-E]} = 0
\earay
for arbitrary variations $\phi_{\perp}$ away from $\phi_0$. The resulting path is then given by the Euler-Lagrange equation
\baray
\label{MPEP}
\frac{\partial^2 \phi_0(x,\lambda)}{\partial \lambda^2} + \nabla^2 \phi_0(x,\lambda) - \frac{\partial V(\phi_0(x,\lambda))}{\partial \phi} = 0\,\,,
\earay
where $\lambda$ is related to $r$ by the relation
\baray
\frac{dr}{d\lambda} = \sqrt{2[U(\lambda)-E]}\,\,.
\earay
Note that (a) $\lambda$ is a real parameter, as we are concerned with the tunneling region where $U > E$, and (b) $\lambda$ is simply a 
parameter we introduced to solve the equation of motion and, in particular has no relation with the Euclidean time.
If a solution to Eq.(\ref{MPEP}) exists, then it can be used as the classical trajectory that describes the most probable escape path
along which tunneling occurs. An infinite dimensional problem thus gets reduced to one described by a single parameter $\lambda$
(or, equally well, by $r$).

One can similarly construct the WKB solutions in the classically allowed region of the field configuration space. The WKB wavefunction
for the classically allowed region is given by
\baray
\label{wkb'}
\Psi(\phi_c(x,s)) = \frac{A}{{(2[U(s)-E])}^{1/4}} \exp \left( \frac{i}{\hbar}\int_0^s ds' \sqrt{2[E-U(s')]} + ...\right)\,\,,
\earay
where the subscript ``$c$" in $\phi_c(x,s)$ means that this classical solution corresponds to a classically allowed trajectory.
Here $s$ parameterizes the path. The classical orbit can be determined by solving the Euler-Lagrange equation
\baray
\label{orbit}
\frac{\partial^2 \phi_c(x,\eta)}{\partial \eta^2} - \nabla^2 \phi_c(x,\eta) + \frac{\partial V(\phi_c(x,\eta))}{\partial \phi} = 0\,\,,
\earay
where $\eta$ is defined as 
\baray
\frac{ds}{d\eta} = \sqrt{2[E - U(\eta)]}\,\,.
\earay

\subsection{Corrections to WKB solutions due to neighboring paths}
Although the main contribution to the WKB approximation comes from the classical path (either Euclidean or Lorentzian),  
there will be additional contributions to it from neighboring paths. In the
case of tunneling, as discussed originally in \cite{Banks:1973ps}, the tunneling takes place in a tubular region that encloses the classical
Euclidean path and also contains various neighboring paths. These neighboring paths give a calculable correction to the WKB solutions.
We shall simply quote the results here. The interested reader should refer to \cite{Bitar:1978vx} for the details. \\

To describe the corrections, one must define coordinates that are orthogonal to the tangent along the classical path in the field configuration
space. These orthogonal coordinates $n(y)$ will be used to describe the nearby paths. Given a classical path $\phi_0(x,r)$, one
can describe a nearby path $\phi(x,r)$ as (from Eq.(\ref{perp}))
\baray
\phi(x,r) = \phi_0(x,r) + \Delta \phi_{\perp}(x,r)\,\,,
\earay
where the perpendicular displacement (with respect to the $\phi_0(x,r)$) is given by
\baray
\Delta \phi_{\perp}(x,r) = \int d^3y \hat{\phi}_{n(y)}(x,r) n(y)\,\,,
\earay
with $\hat{\phi}_{n(y)}(x,r)$ being the unit vector in the orthogonal direction $n(y)$.  Here $y$ describes the coordinates along the orthogonal direction. 
Thus infinitesimal variations of an arbitrary path $\phi(x,r)$ at constant value of $r$ are given by
\baray
\delta \phi(x) |_r = \hat{\phi}_{n(y)}(x,r) \delta n(y)\,\,.  
\earay
Further, let us define the second variations of the action in the classically forbidden and allowed regions as respectively 
\baray
\kappa(y, y' ; r) = \frac{1}{i} \int \int d^3x_1 d^3x_2  \frac{\delta^2 S_{(0)}(\phi)}{\delta \phi_{\perp}(x_1) \delta \phi_{\perp}(x_2)} \mid_{\phi_0(x,r)} \times \hat{\phi}_{n(y)}(x_1,r) \hat{\phi}_{n(y')}(x_2,r), \nonumber \\
K(y, y' ; s) = \int \int d^3x_1 d^3x_2  \frac{\delta^2 S_{(0)}(\phi)}{\delta \phi_{\perp}(x_1) \delta \phi_{\perp}(x_2)} \mid_{\phi_c(x,s)} \times \hat{\phi}_{n(y)}(x_1,s) \hat{\phi}_{n(y')}(x_2,s)\,\,.
\earay
Including the effect of neighboring paths, the wavefunctional for the classically forbidden region around the classical tunneling path is given by
\baray
\label{forb}
\Psi(\phi(x,r)) = \frac{A}{{(2[U(r)-E])}^{1/4}} \exp \left( -\frac{1}{\hbar}\left[\int_0^r dr' \sqrt{2[U(r')-E]} + \right. \right. \nonumber \\
\left. \left. \frac{1}{2}\int d^3y \int d^3y'
\kappa (y, y',r) \delta n(y) \delta n(y') +...\right]\right)\,\,.
\earay
Note that the paths away from the classical path give exponentially suppressed contributions, and this justifies the leading order expectation
that tunneling takes place in a tubular region around the classical solution.
Similarly, the wavefunctional for the classically allowed region around the classical orbit is given by
\baray
\label{allow}
\Psi(\phi(x,s)) = \frac{A}{{(2[E - U(s)])}^{1/4}} \exp \left( \frac{i}{\hbar}\left[\int_0^s ds' \sqrt{2[E -U(s')]} +  \right. \right. \nonumber \\
\left. \left. \frac{1}{2}\int d^3y \int d^3y'
 K(y, y',s) \delta n(y) \delta n(y') +...\right]\right)\,\,.
\earay
The above WKB solutions can be written in a slightly different form which serve better for studying the matching conditions c.f. Sec.(\ref{ss:matching}).  One defines
a ``mass" parameter as
\baray
\label{mass}
m(\lambda) = \int d^3x \left( \frac{\partial \phi}{\partial \lambda} \right)^2 \,\,.
\earay
Note that the mass parameter is always positive, $m(\lambda) > 0$. Note that from Eq.(\ref{path}), $dr/d\lambda = \sqrt{m(\lambda)}$. 
Now one can define the WKB solutions both in the tunneling region and the
allowed region via the same parameter $\lambda$ as 
\baray
\label{forb'}
\Psi(\phi(x,\lambda)) = \frac{A}{{(2[U(\lambda)-E])}^{1/4}} \exp \left( -\frac{1}{\hbar}\left[\int_{0}^{\lambda} d\lambda' \sqrt{2m(\lambda')[U(\lambda')-E]} + \right. \right. \nonumber \\
\left. \left. \frac{1}{2}\int d^3y \int d^3y' \kappa (y, y',r) \delta n(y) \delta n(y') +...\right]\right)
\earay
in the forbidden region and
\baray
\label{allow'}
\Psi(\phi(x,\lambda)) = \frac{A}{{(2[E - U(\lambda)])}^{1/4}} \exp \left( \frac{i}{\hbar}\left[\int_{0}^{\lambda} d\lambda' \sqrt{2m(\lambda')[E -U(\lambda')]} + \right. \right. \nonumber \\
\left. \left. \frac{1}{2}\int d^3y \int d^3y' K(y, y',s) \delta n(y) \delta n(y') +...\right]\right)
\earay
in the allowed region, respectively.

\subsection{Matching conditions} \label{ss:matching}

The greatest simplification that has happened in the above analysis is that tunneling in the infinite dimensional configuration space
has been reduced, to the leading order, to a quantum mechanics problem in one dimension. The effect of nearby paths appears first
at next to leading order. Let us first consider the problem of tunneling across one barrier, and then the case of tunneling across 
two barriers (and the condition for resonance tunneling):

\begin{figure}
\begin{center}
\includegraphics[width=9cm]{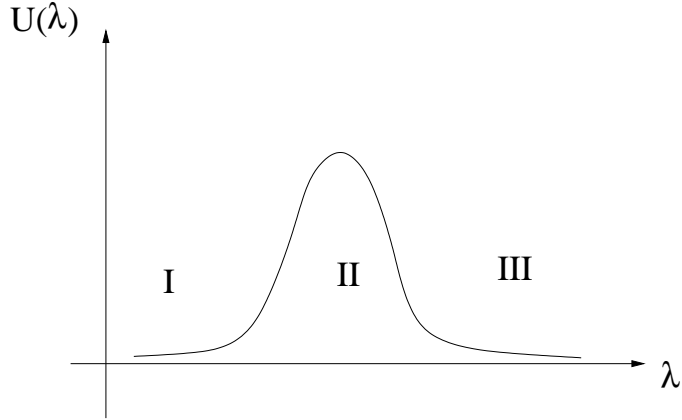}
\vspace{0.1in}
\caption{The potential $U(\lambda)$ as a function of the parameter $\lambda$.}  
\label{fig:U123}
\end{center}
\end{figure}

\begin{itemize}
\item {\bf Tunneling across a single barrier}: 
As in Fig.(\ref{fig:U123}), consider three regions, regions I, II and III. The energy $E$ is such that regions I and III are classically allowed
whereas region II is classically forbidden. As in \cite{Tye:2006tg}, let $\alpha_R$ and $\alpha_L$ be the coefficients of the right
-moving and the left-moving components, respectively, in region I.  Similarly, let $\beta_R$ and $\beta_L$ be the coefficients 
in region III. The two turning points are given by $\lambda_1$ and $\lambda_2$. Then the WKB connection formulae imply
\baray
\left(
\begin{array}{c}
\alpha_R \\
\alpha_L \end{array}
\right) 
=
\frac{1}{2}\left(
\begin{array}{cc}
\Theta + \Theta^{-1}    & i(\Theta - \Theta^{-1}) \\
-i(\Theta - \Theta^{-1})& \Theta + \Theta^{-1} \end{array}
\right)
\left(
\begin{array}{c}
\beta_R \\
\beta_L \end{array}
\right) \,\,,
\earay 
where $\Theta$ is given by
\baray
\label{theta}
\Theta \simeq 2 \exp \left( \frac{1}{\hbar}\left[\int_{\lambda_1}^{\lambda_2} d\lambda' \sqrt{2m(\lambda')[U_{eff}(\lambda')-E]} \right] \right)\,\,,
\earay
and where $U_{eff}(\lambda)$ is the result of integrating out the fluctuations (denoted in the previous section by $\delta n$) around the tunneling path (i.e., the Euclidean classical solution).  $U_{eff}(\lambda) = U(\lambda) + F(\lambda)$, where
$F(\lambda)$ contains the one-loop effect that appears when one integrates out the fluctuations denoted by $\delta n$. To evaluate $F$ we use  
\baray
\label{F}
\exp \left(-\frac{1}{\hbar}\int_{\tau_1}^{\tau}F(\tau')d\tau' \right) = \left[ det' \left(-\frac{\partial^2}{\partial \tau^2}-
\nabla^2 + \frac{\partial^2 V(\phi)}{\partial \phi^2} \right)  \right]^{-1/2}\,\,,
\earay 
where the prime on $det'$ implies, as usual, that the zero mode is to be excluded while evaluating the determinant.
The variable $\tau$ is related to $\lambda$ by $\frac{d\tau}{d\lambda} = \sqrt{\frac{m(\lambda)}{2[U(\lambda)-E]}}$, 
and we write $\tau_1 = \tau(\lambda_1)$ etc.  The difference between the simple WKB tunneling formula in quantum mechanics and the one
in quantum field theory is the inclusion of the gradient term $\frac{1}{2}(\nabla \phi)^2$ and the one-loop effect $F$.
The transmission probability from region I to region III is then given by
\baray
T_{I \to III} = |\frac{\beta_R}{\alpha_R}|^2 = 4(\Theta + \frac{1}{\Theta})^{-2} \simeq \frac{4}{\Theta^2} \,\,.
\earay

\item {\bf Resonance Tunneling}:
We consider the tunneling across two barriers, Fig(\ref{fig:U12345}). Regions I, III, and V are classically allowed, whereas regions II and IV 
are classically forbidden. Again, one can follow through the analysis of \cite{Tye:2006tg}, keeping in mind that one should also
include the gradient term and the one-loop effect. Let $\gamma_R$ and $\gamma_L$ be the right-moving and the left-moving 
coefficients in the region V. 

\begin{figure}
\begin{center}
\includegraphics[width=9cm]{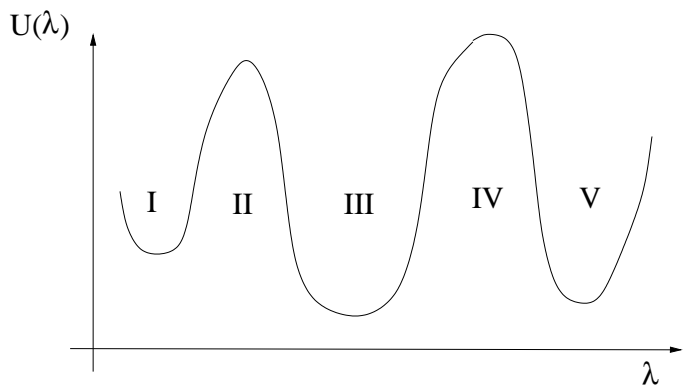}
\vspace{0.1in}
\caption{A potential with multiple barriers and minima. Resonance tunneling can happen as a result of constructive interference.} 
\label{fig:U12345}
\end{center}
\end{figure}

Then we have   
\baray
\left(
\begin{array}{c}
\alpha_R \\
\alpha_L \end{array}
\right) 
=
\frac{1}{2}\left(
\begin{array}{cc}
\Theta + \Theta^{-1}    & i(\Theta - \Theta^{-1}) \\
-i(\Theta - \Theta^{-1})& \Theta + \Theta^{-1} \end{array}
\right) \times
\left(
\begin{array}{cc}
e^{-\frac{i}{\hbar}W} & o \\
0                     & e^{\frac{i}{\hbar}W} \end{array}
\right) \nonumber \\
\times
\left(
\begin{array}{cc}
\Phi + \Phi^{-1}    & i(\Phi - \Phi^{-1}) \\
-i(\Phi - \Phi^{-1})& \Phi + \Phi^{-1} \end{array}
\right) \times
\left(
\begin{array}{c}
\gamma_R \\
\gamma_L \end{array}
\right) \,\,,
\earay
where
\baray
\label{phi2}  
\Phi \simeq 2 \exp \left( \frac{1}{\hbar}\left[\int_{\lambda_3}^{\lambda_4} d\lambda' \sqrt{2m(\lambda')[U_{eff}(\lambda')-E]} \right] \right) \,\,,
\earay
and
\baray
\label{W2}  
W = \int_{\lambda_2}^{\lambda_3} d\lambda' \sqrt{2m(\lambda')[E - U_{eff}(\lambda')]} \,\,.
\earay
The transmission probability from region I to region V via region III is now given by
\baray
\label{ItoV}
T_{I \to V} = 4 \left( \left( \Theta \Phi + \frac{1}{\Theta \Phi}\right)^2 \cos ^2 W + \left( \frac{\Theta}{\Phi} + \frac{\Phi}{\Theta} \right)^2  \sin^2 W \right)^{-1}\,\,.
\earay
Again, although this resembles the quantum mechanical result in Sec.(\ref{QM}), the crucial difference in field theory is that
one must consider the effective potential $U_{eff}(\phi)$ in all the WKB results and not the potential $V(\phi)$.
Now one can easily see the resonance condition
\baray
\label{reson2} 
W = \left( n + \frac{1}{2} \right) \pi \,\,,
\earay
where $n = 0,1,2,3,...$.
\end{itemize}

\section{Branes and Tunneling} \label{dbi}
Here we discuss the role branes play in vacuum transitions, necessitated by their central role in realistic backgrounds in flux compactifications.
The primary ingredient in stable string compactifications are wrapped branes, and the charges
they impart on the various cycles of the

internal
 geometry.  A subset of these branes are
mobile in the sense that they are point-like in the six-dimensional bulk.  Both the point-like and
wrapped branes contribute a significant fraction of the
 landscape\cite{Gomis:2005wc,DeWolfe:2007hd}, although their dynamics is quite different than the closed string moduli which control the topological and geometric data of the 
 vacua\footnote{In F-theory,
 however, the closed and the open string landscape can be treated in a unified manner.}. The brane dynamics is decribed by the DBI action which has a nonstandard kinetic
term. Decay processes involving the brane, therefore, are described by DBI tunneling which, as we shall see, can differ significantly from
the Coleman-De Luccia tunneling \cite{Brown:2007ce}.
Before describing DBI tunneling, we  will briefly explain other vacuum transition processes involving branes.  

The Brown-Teitelboim instanton \cite{Brown:1987dd} describes a tunneling event whereby a volume-form flux is changed discretely.  Charge conservation dictates that the volume form be discontinuous across the domain wall by an amount equal to the brane charge, $\mu$.  The domain wall in this case is the charged brane itself, and its tension is $\tau$.  Because the brane is of zero thickness, the decay
rate through such a process is exactly described by the thin wall limit of Coleman-De Luccia.
The volume-form flux which couples to the brane contributes an amount 
\be
V = \frac{1}{2}\ast \left( F_{4} \wedge \ast F_{4}\right)
\ee
to the vacuum energy, and so the energy difference between the two vacua is given by
\be
\epsilon =  \mu\sqrt{2 V_-} - \frac{1}{2}\mu^2\,\, ,
\ee
where $V_-$ is the contribution to the true vacuum energy due to $F_{4}$. 

A more exotic form of brane-induced vacuum decay is the Kachru-Pearson-Verlinde instanton \cite{Kachru:2002gs}.  In this scenario, a flux stabilized, ``doubly contractible" spherical brane tunnels by slipping around a larger sphere, and the potential barrier height is given by the brane tension times its 
spatial
 volume.  As expected, a higher potential causes greater suppression of the decay rate.

\subsection{DBI tunneling}
DBI tunneling is distinct from the previous two scenarios in that no quantum numbers need be changed. 
Instead, a single mobile brane simply tunnels between meta-stable loci within the 
internal
geometry.  What is unusual about this instanton, is that the decay rate increases when the potential barrier height is elevated to near the (five dimensional) BT threshold.  This effect is due to the decreasing bubble wall tension, to which the decay rate is exponentially sensitive.  To illustrate this, we consider a mobile D3 (or anti-D3) brane in a warped bulk geometery.  Here and below we follow \cite{Brown:2007ce} closely.  The action for the brane in a warped background is given by \footnote{The essential features of DBI tunneling do not depend on the warp factor.} 
\be 
S =  \int  d^4x \left(\frac{1-\sqrt{1 - f(\phi) {\partial_\mu\phi \partial^\mu \phi}} }{f(\phi)}- V(\phi) \right) \,\, ,
\ee
where $1/f(\phi)$ is the warped brane tension. 
The potential $V(\phi)$ has two important sources:  (1) any five-form flux component which is incompatible
with the supersymmetry preserved by the brane, and (2)
threshold corrections from K\"ahler stabilizing D7 branes.

We would like to compute the rate per unit volume for the three-brane to transition between
two local minima of the potential $V(\phi)$.  The rate for this process is given by the path integral
over the Euclidean action
\ba
\Gamma/V &=& \int[D\phi] e^{-S_E[\phi]} \\
&\sim& e^{-S_E[\phi_{cl}]-S_E[\phi_+]} \,\, .
\ea
We are using the  semi-classical approximation, where the field corresponding to the
classical Euclidean solution $\phi_{cl}$ dominates the path integral.  Proper normalization
is achieved by subtracting the Euclidean action of the constant field in the false
vacuum, and we assume the fluctuation determinant sets the overall coefficient to order one in fundamental units.  
We call the field $\phi_{cl}$ ``the bounce," and it is the unique non-constant solution to the equations of motion  with lowest Euclidean action which asymptotically approaches the false vacuum at Euclidean infinity.  Such a solution has been shown\cite{Coleman:1977th} (under reasonable assumptions) to be a maximally symmetric interpolating solution, i.e. its location breaks four generators, but preserves the remaining $O(4)$ symmetry of the underlying (Wick rotated) Poincar\'e group.  This allows us to write the Euclidean action as
\be \label{eq:ignore}
S_E = 2 \pi^2 \int \rho^3 d\rho \left(\frac{1}{f(\phi)}\sqrt{1 + f(\phi) \dot{\phi}^2} -\frac{1}{f(\phi)}+ V(\phi) \right) \,\, ,
\ee
where dots represent derivatives with respect to $\rho$.

We focus on two neighboring minima of $V(\phi)$, e.g. A and B of Fig. ($1$).  A vacuum transition is much like a first order phase transition, and begins with the formation of a bubble of true vacuum within the false vacuum.  The bubbles will tunnel quantum mechanically to a critical radius $\bar{\rho}$, and then expand classically at essentially the speed of light, converting the vacuum energy difference into kinetic energy of the bubble wall.
The high degree of symmetry has reduced our computation to the one-dimensional motion of a particle in a potential $-V(\phi)$, albeit with an unusual kinetic term.  In the language of Sec.(\ref{QFT}), the solution to this one dimensional equation of motion gives the most probably escape path
(MPEP). 
In our analysis, we will use conjugate momentum, $\pi_\phi$, and Hamiltonian, $H$, which are defined with respect to $L^{\rm canonical}/(2\pi^2 \rho^3)$.
\ba
L = \frac{L^{\rm{canonical}}}{2\pi^2\rho^3} &=& \frac{1-\gamma}{f \gamma} + V \,\, , \\ 
\pi_\phi = \frac{\partial L}{\partial \fid}  &=& \fid \gamma \,\, , \\
H = \pi_\phi \fid - L  &=& \frac{1-\gamma}{f} - V\label{eqn:hamiltonian} \,\, , 
\ea
where
\ba
\gamma^2 = \frac{1}{1 + f(\phi)\fid^2} = 1 - f(\phi) \pi_\phi^{2} \label{gammadef}\,\, . 
\ea

As $\pi_\phi \to f(\phi)^{-1/2}$ the field velocity $\fid \to \pm\infty$.  This means that when one plots
the instanton in $\rho$-$\phi$ coordinates, $\phi$ is not a single valued function of $\rho$. This failure of coordinates
is due to the fact that the world line in Euclidean space continues to curve in response
to the forces imparted by the potentials, and $\rho$ need not be always increasing along
the world line. One encounters a similar situation with the $S^1$ instanton that describes 
Schwinger pair production in the presence of an external electromagnetic field where the
Euclidean classical solution is double valued as a function of the Euclidean time. This,
of course, is not surprising as the relativistic action for a charged particle in an
external electromagnetic field is the one dimensional version of the DBI action that we
are considering here. The Euclidean equation of motion is given by
\ba
\label{dbibounce}
\dot{\pi}_\phi = -\frac{\partial H}{\partial \phi} - 3 \frac{\pi_\phi}{\rho} \label{eqn:pidot} \,\, .
\ea

Notice that
\be
\label{eqn:hdot}
d H = - \frac{3}{\rho} \pi_\phi d\phi 
\ee
which determines the rate of non-conservation of $H$.

Smoothness of the brane at $\rho = 0$ requires that $\phi$ begin (and end) with zero velocity $(\gamma = 1)$, so $dH$ integrates to the energy difference between the two vacua.
A fraction of this energy is matter inside thick-walled bubbles.  For thin-walled bubbles, the energy is purely vacuum energy, causing the well known ``empty universe problem."

Thin walled bubbles occur whenever $S_1 \gg \epsilon/\mu$, where $\epsilon$ is the difference in vacuum energy, $\mu$ is the mass of $\phi$ (in either vacuum), and $S_1$ is the tension of the domain wall.  Then we can consider the energy nonconservation to be negligible (i.e. $dH = 0$ upto 
$\mathcal{O}(\epsilon)$ in Eq.(\ref{eqn:hdot})). The mass $\mu$ sets the scale for the bubble wall thickness, and the ratio $S_1/\epsilon$ determines the bubble radius c.f. Eqn.(\ref{eq:rhobar}).  In the inverted potential, $\phi$ sits exponentially close to the true vacuum, $\phi_{-}$, for a long time ($\rho \gg 1/\mu$) before rolling off and continuing quickly to the top of the false vacuum, $\phi_{+}$.  The thin wall bounce is approximately a step function.  A first integral of the equation of motion Eq.(\ref{dbibounce}) is given by writing $H + \mathcal{O}(\epsilon) = E$ in Eq.(\ref{eqn:hamiltonian}), yielding
\be
\gamma(\phi,\pi_\phi) = 1 - f(\phi)\left(V(\phi) +E + \mathcal{O}(\epsilon)\right) \label{eqn:solvegamma} \,\, , 
\ee
where $E = -V_+$, the value of the potential at the false vacuum.  This is solved for the momentum to give
\be
\pi_\phi = \sqrt{V_0(\phi)\left(2 - f(\phi) V_0(\phi)\right)}  \label{eq:piofphi}\,\, , 
\ee
where we have absorbed $E + \mathcal{O}(\epsilon)$ into the potential by setting {\em both} minima of $V_0$ equal to zero.  
The only obstacle to modeling friction as a conservative force is that it is momentum dependent rather than position dependent.  But if the momentum can be shown to be purely a function of position, as is done with Eq.(\ref{eq:piofphi}), the obstacle disappears, and we may write an effective (but  trajectory-dependent) potential which completely describes the forces, and thus motion.   Defining $V_0$ is simply absorbing the friction into the potential, a feat possible because the bounce $\phi(\rho)$ is a monotonic map (in the sense that its inverse is a well defined function).    
The term of order $\epsilon$ is linear in $\phi$ to first approximation, and so we can write
\be
V_0(\phi) = V(\phi) - V(\phi_+) + \epsilon\frac{\phi - \phi_+}{\Delta\phi}    
\ee
From this we can find the solution by inverting the function
\be
\rho(\phi) = \int \frac{1 - f(\phi)V_0(\phi)}{\sqrt{V_0(\phi)\left(2 - f(\phi) V_0(\phi)\right)}}d\phi  \,\, . \label{eq:twsoln}
\ee
A refinement of $V_0(\phi)$ can be made iteratively by solving for $H(\phi)$ using this solution, and plugging it into Eq.(\ref{eqn:solvegamma}). 
 
These solutions determine the bubble nucleation rate through the difference in Euclidean action
$B = S_E[\phi] - S_E[\phi_+]$.  Notice only the interior volume and bubble wall contribute to $B$.

\ba
B &=&  -2 \pi^2 \int_0^{\bar{\rho}} \rho^3   \epsilon d\rho  + 2\pi^2 {\bar{\rho}}^3\int_{\phi^-}^{\phi^+}\sqrt{V_0(2 - f  V_0)}d\phi  \,\, , \nonumber\\
&=& -\frac{\pi^2{\bar{\rho}}^4}{2}\epsilon + 2\pi^2{\bar{\rho}}^3S_1 \,\, ,
\ea
where $\bar{\rho}$ is the radius of the bubble.
We may integrate Eq.(\ref{eqn:hdot}) to solve for
\be
\bar{\rho} = 3 S_1/\epsilon\,\,.\label{eq:rhobar}
\ee
The tunneling rate per unit volume is given by
\be
\label{wrink}
\Gamma/ V \sim e^{-27\pi^2 S_1^4/2\epsilon^3} \,\, . 
\ee
The bubble wall tension is $S_1 = \int d\phi\sqrt{V_0(2-fV_0)}$, so the tunneling rate is large when $f V_0 \to 2$ over much of the potential.  This tunneling rate is always greater than the Coleman - De Luccia rate in the same potential. Although in general this does not imply that the tunneling is rapid, but if $fV_0$ is close to $2$, the rate can be substantially greater than Coleman-De Luccia.

 \subsection{The Wrinkle}
As one might expect from a Euclidean solution, there is a generic multi-valuedness  
of $\phi(\rho)$.  The complete range of $\gamma$ is given by
\be
-1 \leq \gamma \leq 1 \quad\Longrightarrow \quad 0 \leq f(\phi) V_0(\phi) \leq 2 \,\, .
\ee

\begin{figure}[h] 
   \begin{center}
   \includegraphics[width=9cm]{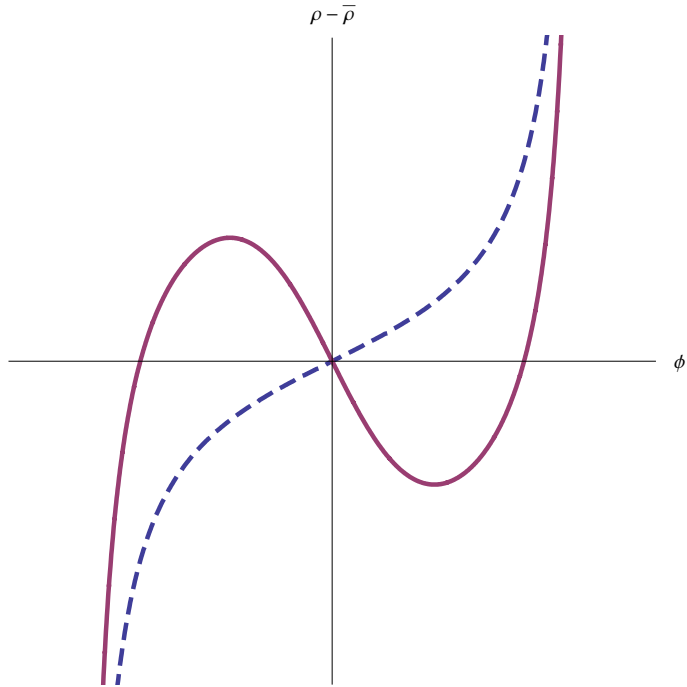} 
   \caption{Two solutions for the instanton trajectory, (dashed) for a low barrier ($fV_0 < 1$) and (solid) for a high barrier ($1<fV_0<2$).} 
   \label{profile}
   \end{center}
\end{figure}

A typical solution would look something like Fig.(\ref{profile}).
It is clear why the solutions for large potentials is ``wrinkled," i.e. folds over and becomes multivalued.  This is because the extrinsic curvature of the solution is determined by the gradient of the potential.  This is the Euclidean version of $F = m a$. While finite acceleration prevents an initially time-like world line from wrinkling, no such effect occurs in Euclidean space (as is evident from the $S^1$ instanton that exists
for the Schwinger pair production for charged particles in external electromagnetic field).

One might worry that the presence of the wrinkle (and the consequent large value of the second derivative of $\phi$) implies a breakdown
of the DBI description of the brane. However, the correct measure of how big the $\alpha'$ corrections are is the extrinsic curvature
of the brane and not just the second derivative of $\phi$ with respect to $\rho$. The DBI action can be trusted whenever curvatures are low
\cite{Leigh:1989jq}. In the thin wall regime, the extrinsic curvature of the solution $\phi(\rho)$ in the warped background is given by the covariant derivative
of a unit one-form $\hat{n}$ normal to the curve $\phi(\rho)$, whose pullback is traced over using the induced metric.
\ba
K(\phi)&=& (\nabla_\mu \hat{n}_\nu)h^{i,j}\partial_i x^\mu \partial_j x^\nu    \nonumber \\
&=& \frac{f(\phi)^{3/4}}{\sqrt{\alpha'}} \frac{\partial}{\partial \phi} \left( V(\phi)- \frac{1}{f(\phi)} \right) \,\, .
\ea
The bulk metric is given by the Euclidean warped throat: $ds^2 = \alpha' f^{-\frac{1}{2}} d\rho^2 + \alpha'^2d\phi^2$.
For the validity of the DBI description $K(\phi)$ should be smaller than the string scale $1/\sqrt{\alpha'}$. This limits how steep the potential can be
while still trusting the DBI description.

\subsection{Including the effects of gravity}
Gravity complicates the calculation in a number of ways.  The main effect is that the background is no longer flat, although we still assume that an $O(4)$ symmetry is preserved in the maximally symmetric, uniformly curved background.  Another effect is that the friction felt by the rolling $\phi$ is now dynamical, and coupled to the inflaton Hamiltonian.  Although this coupling makes the general solution rather difficult to approximate, the thin wall limit remains explicitly calculable.  Here we simply quote the results \cite{Brown:2007ce}.

The decay rate in the presence of gravity (with $M_P$ the reduced Planck mass) is given by
\be
\Gamma/V \sim e^{-\left(B_{\rm wall} + B_{\rm inside}\right)}\,\, ,
\ee

where
\ba
B_{\rm inside} &=& \left\{\frac{-12\pi^2 M_P^4}{V}\left[1 - \left(1 - \frac{\bar{\rho}^2 V}{3 M_P^2}\right)^\frac{3}{2}\right]\right\}^{V = V_-}_{V = V_+} \,\, ,\\
B_{\rm wall} &=& 2\pi^2\bar{\rho}^3S_1 \label{Bwall}\, \, .
\ea

Here
\be
\bar{\rho} = \frac{3 S_1M_P}{\sqrt{\epsilon^2 M_P^2 + \frac{3}{2} S_1^2 \left(V_+ + V_-\right) +\frac{9}{16} S_1^4/M_P^2 }}\,\,,
\ee

and
\be 
S_1 = \int_{\phi_-}^{\phi_+} d\phi \sqrt{V_0(\phi)(2 - f(\phi)V_0(\phi))} \,\, , \label{gravtension}
\ee 
as before.

\subsection{Efficiency} \label{efficiency}

The Brown-Teitelboim instanton is responsible for neutralizing any potential which exceeds a certain critical threshold analogous to the pair-creation threshold.  One might expect this to lead to a large population of vacua which, despite having high potential barriers, are below the BT threshold.  It is precisely such an ensemble of vacua in which DBI tunneling is much more rapid than the standard CDL result, but still slow enough to be within the semi-classical region of validity.  One would expect that the truly rapid case (i.e. where tunneling is so rapid that the term meta-stability is no longer operative) exists in two distinct limits.  The first is simply the small barrier behavior near a critical point, i.e. where a first order transition nears second order.  The second is where resonant effects become large, causing
the barrier to resemble a reflectionless potential (we shall see in the next section how the DBI behavior can lead to this).  
\footnote{A third, more speculative limit exists where the potential
rises so steeply as to bypass the intermediate-height regime, i.e. a step function.  It should be noted that this
third
limit only exists far outside the range of validity of the DBI action, and we would have to assume that the unsuppressed  $\alpha'$ corrections do not conspire to prevent these rapid transitions.  We will not consider this possibility here.}

\subsection{Resonant DBI and the Moonscape} \label{moon}
\begin{figure}[h] 
   \begin{center}
   \includegraphics[width=10cm]{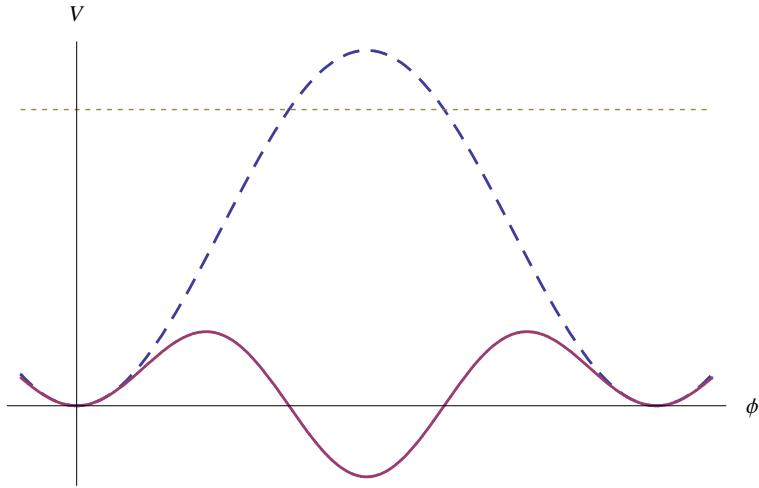} 
   \caption{The dashed curve represents the actual potential energy $V_0$ for a supercritical barrier, and the solid curve is the effective (non-relativistic) potential $V_0^{CDL}$ seen by the field.  The many craters appearing from high open string barriers make up ``the moonscape."}
   \label{fig:VCDL}
   \end{center}
\end{figure}

Two interesting features appear when DBI tunneling is combined with super-critical potential barriers, i.e. those for which $f V_0 > 2$ over
any part of the barrier.  The first is that the BT mechanism allows for nucleation of D3 bubbles of various topologies.  For relatively low barriers, the only allowed topology consistent with our $O(4)$ invariant ansatz is $S^1\times S^3$, but for higher potentials an $S^4$ bubble will be dominant.  (The $S^4$ requires a higher potential since it must occur near $\rho = 0$, where friction is large.)
The second, and more striking feature is that mobile D3's will see the barrier as effectively {\em two separate barriers} of equal height (see Fig.(\ref{fig:VCDL})), since the interior supercritical region is classically allowed.  This doubling of vacua naturally leads to a setting where resonance conditions can potentially be satisfied.  It is this possibility which enables the transition to be much more dominant than the closed D3 bubble solutions, although Euclidean methods cannot show this.  
Additionally, this provides a mechanism for {\em pairwise similar} barriers, causing a statistical enhancement for the resonance condition to be satisfied.  The resulting picture is that the effective maximum barrier height corresponds to $V_0 = 1/f$, since all higher potentials contribute less to the domain wall tension density.  The reduced tension can be described by the CDL-effective potential 
\be
V_0^{CDL} = V_0 (1 - f V_0/2) \label{eq:vcdl}\,\,,
\ee
where the CDL (canonical) kinetic term is assumed.
In the following subsection we will show that the correct way to calculate the domain wall tension, and thus nucleation rate for supercritical barriers will be to use the above effective potential $V_0^{CDL}$ with a standard kinetic term, via the methods of section \ref{resonance}. 
Thus high `mountains' in the open string landscape are turned into `craters,' since the only suppression occurs in the window $0 < f V_0 < 2$.  The open string landscape is really a moonscape.  Because the supercritical barrier regions contribute nothing to the domain wall tension, long distances in field space are just as accessible as short ones, a criteria which can lead to {\em percolation}.

\subsection{Craters and resonance tunneling} \label{KG}
As we have just shown, the DBI action leads to an enhanced tunneling rate which can be
traced to the fact that the domain wall tension is corrected to $S_1 = \int d\phi \sqrt{V_0 (2 - f V_0)}$.
Equivalently, the potential can be reduced to a CDL-effective potential given by Eq.(\ref{eq:vcdl}).
In fact, this equivalence is not just limited to D-branes.
In this section, we show how this equivalence can be easily seen in the case of a point particle obeying the
Klein-Gordon equation.
 \begin{figure}[h] 
   \begin{center}
   \includegraphics[width=9cm]{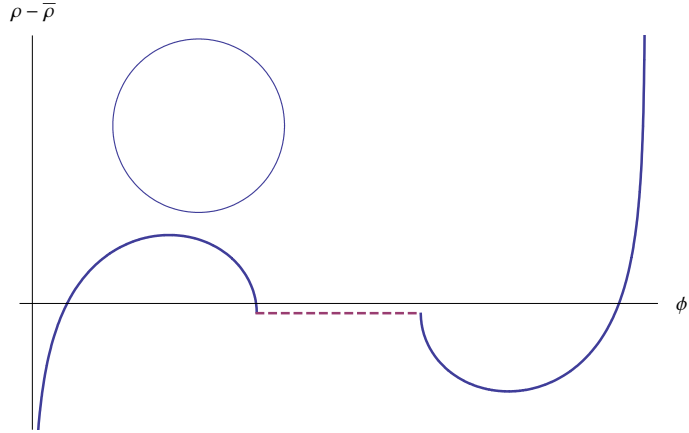} 
   \caption{Two solutions for the supercritical $(f V_0 > 2)$ case, an $S^1\times S^3$ bubble, and the ``Klein-Tunneling" solution with a potentially resonant interior.  The solutions differ only by initial conditions, since the transition must begin and end in the two vacua.
   The $S^1\times S^3$ D3 bubble dominates over the D3 transition, unless the resonance condition is satisfied.
   The $S^4$ D3 bubble occurs far below, near $\rho = 0$.}
   \label{fig:klein}
   \end{center}
\end{figure}
For moderately high barriers, the effective potential resembles a crater, and when the barrier height
exceeds the BT threshold, the crater interior will be a classically allowed region, as in Fig.(\ref{fig:VCDL}).  As pointed out in \cite{Brown:2007ce}, it is in such cases where no interpolating Euclidean solutions exist, although we expect
the transition rate to be non-zero.  This has been referred to as `Klein tunneling'\cite{Calogeracos:1998rf}, which we illustrate in Fig.(\ref{fig:klein}).  In the absence of resonant effects, the closed (BT) instanton is dominant, but satisfying the resonance condition will make the tunneling solution greatly favored.
To calculate the transition rate, we will resort to a one-dimensional
example of large barrier penetration by solving the Klein-Gordon wave equation.

Consider a particle of unit charge and mass $m$ interacting with an electromagnetic potential.  For simplicity, we assume no warping.  The action is given by
\be
S = -\frac{1}{2}\int dx dt \left({\overline {D_\mu \Psi}} D^\mu \Psi + m^2 \overline{\Psi}\Psi\right)\,\,,
\ee
where 
\baray
A_\mu = \left(
\begin{array}{c}
V\\
0
\end{array}\right)_\mu ,\,\,\,\,\,\,\,
D_\mu = \partial_\mu - i A_\mu\,\,.
\earay

The resulting Klein-Gordon equation $D_\mu D^\mu \Psi = m^2 \Psi$ can be solved using separation of variables.  Here we consider
a single positive frequency mode of energy $\omega$.  The choice $\Psi = e^{i \omega t} \psi(x)$ removes the gauge symmetry $A_t \to A_t + \partial_t \alpha$, and so we shall replace $V$ with the gauge fixed $V_0$.  The time independent Klein-Gordon equation is then
\be
-\frac{1}{2 \omega} \frac{\partial^2}{\partial x^2} \psi + V_0\left(1 - \frac{V_0}{2\omega}\right)\psi = \left(\frac{\omega}{2} - \frac{m^2}{2\omega}\right)\psi \,\,.
\ee
We have written this equation in a form which allows for manifest mapping of the Klein-Gordon equation to the Schr\"odinger equation.  In so doing we have defined the Schr\"odinger effective potential $V^{Sch}_0 = V_0\left(1 - V_0/(2 \omega)\right)$ in complete analogy with Eqn.(\ref{eq:vcdl}).  It is now clear how we may use the non-relativistic formalism to describe relativistic resonance tunneling.
In particular, DBI tunneling through a single barrier will be sensitive to resonant effects whenever the barrier is higher than twice the local D3 tension.

\section{Percolation in the Landscape: Efficient Tunneling and Preferred Paths} \label{percolate}

The interesting consequence of both resonance and DBI tunneling is that when certain conditions are met, tunneling rates become highly enhanced. 
This effect can be traced to a reduced domain wall tension.
In the case of resonance tunneling, tunneling becomes insensitive
to the height and the width of the barrier (provided the resonance condition is exactly satisfied), and  tunneling proceeds ``classically," i.e. without any suppression.
The time scale of tunneling will be the natural scale of the problem. In cosmological scenarios, this time scale will the Hubble
time. Similarly, in DBI tunneling if the right conditions are met (i.e. $1< fV_0$), tunneling becomes much more efficient.
This is exponentially faster than the time scale afforded by usual tunneling, although still meta-stable in the controlled regime.
When $f V_0 > 2$, DBI tunneling naturally allows for resonant tunneling (Sec.(\ref{efficiency}) and Sec.(\ref{moon})). 
In fact, as we later explain in Sec. \ref{cheby}, even in the absence of any resonance / DBI tunneling, the enormous number of decay
channels expected in a flux landscape can make vacuum decay a rapid phenomenon if certain conditions are met.
Such efficient tunneling will 
have nontrivial consequences for cosmology, especially in the 
context of the string landscape which provides a huge number of vacuum sites and decay channels, and where the DBI action describes 
a large fraction of the transitions. In \cite{Tye:2006tg}, Tye has suggested there might be a possible solution to the cosmological constant problem using such efficient tunneling, \cite{Davoudiasl:2006ax} uses efficient tunneling to extract 
some interesting domain wall phenomenology from the last few e-folds of inflation (discussed in Sec.(\ref{discuss})). The crucial question, of course, is how 
frequently the conditions conducive to efficient tunneling
are satisfied in the landscape. If it is extremely rare to find conditions for efficient tunneling, then one might question the efficacy of efficient
tunneling. In particular, mere isolated instances of efficient tunneling in the landscape will not alter the overall tunneling dynamics across
the landscape.\\

In this section we shall argue, along the lines of \cite{Tye:2006tg}, that even if the relative probability of efficient tunneling
is low, the string landscape might have the right characteristics to make use of it. Two facts will be crucial for this
: (i) a large number of field (or flux) directions present in the string landscape that lead to a large number of decay channels, and (ii)
 the possibility to resonance tunnel between  
somewhat far away sites of the landscape (and, as discussed in Sec.(\ref{efficiency}) and Sec.(\ref{moon}), DBI tunneling can itself become
resonant).
The second condition is not entirely necessary. We explain in Sec. \ref{cheby} how even usual vacuum decay can become rapid in
the landscape provided certain conditions are satisfied.
To quantify our discussion, we shall use some basic results from Percolation Theory \cite{Stauffer} and
explore the possibility of finding paths across the landscape that are continuously connected by channels of efficient tunneling. Such paths
will make the landscape {\it percolate}. \\

\begin{figure}
\begin{center}
\includegraphics[width=9cm]{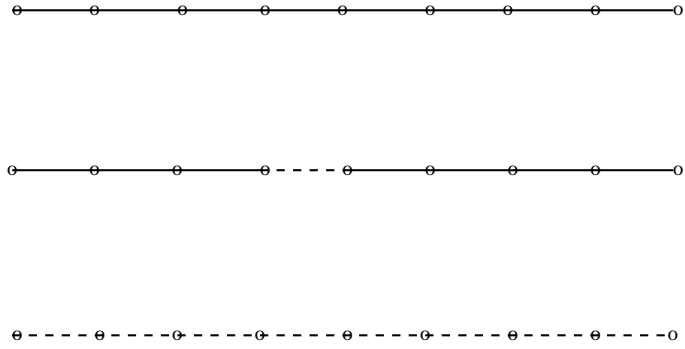}
\vspace{0.1in}
\caption{The small circles represent lattice sites.
Solid lines are paths of infinite resistance. Dashed segments
offer zero resistance. For conduction, all the segments should have
zero resistance, i.e. critical probability $= 1$.}
\label{fig:1d}
\end{center}
\end{figure}

The basic idea underlying percolation theory, and its relevance to the landscape and efficient tunneling, can be illustrated by a simple
example. Consider a two-dimensional grid of electrical resistors. The lattice points are the points where four resistors join 
(see Fig(\ref{fig:1d})). Let this be a large enough grid so that the linear size of the grid is much larger than the linear size of a unit cell of the lattice. 
In other
words, the number of resistors involved is a large number $N$. Considering any two neighboring points on this lattice, let $p$ be the  
probability of the resistance between the two points being zero, 
 and so $1-p$ is the probability that it is infinite.  
The question we ask is : starting from
an initial value of $p = 0$ (i.e., all resistors nonconducting), we slowly increase the value of $p$ (i.e. randomly convert some
of the infinite resistors to zero resistance so that it can perfectly conduct electricity). At what {\it critical} value of $p$,
the critical probability $p_c$, does the grid become conducting?

\begin{figure}
\begin{center}
\includegraphics[width=9cm]{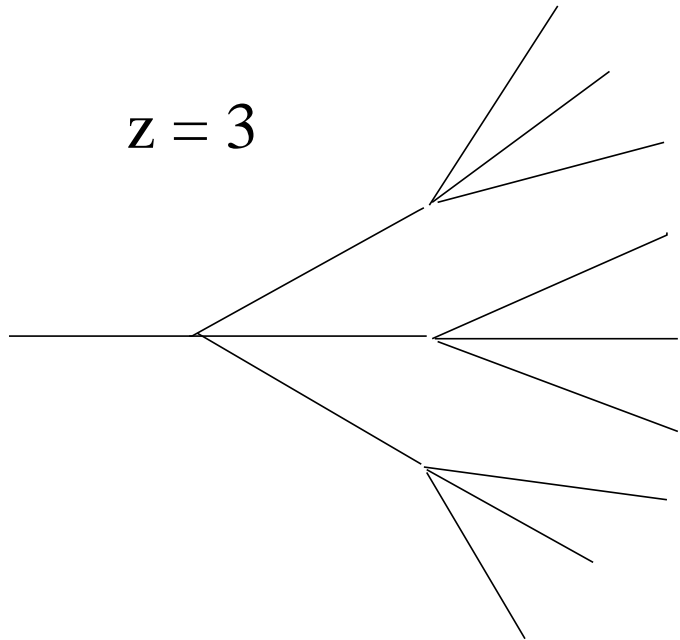}
\vspace{0.1in}
\caption{Bethe lattice with the number of connections to any lattice site given by $z = 3$.}
\label{fig:bethe3}
\end{center}
\end{figure}

At this critical value, there will appear one continuous
connected line of zero resistance 
resistors that will let the current pass through. The two dimensional grid will separate into two non conducting regions,
separated by a conducting line. In other words, the grid will {\it percolate}. 
For two-dimensions, it is known that $p_c = 1/2$ \cite{Stauffer}. Of course, for one-dimension, 
$p_c = 1$ (i.e., all the resistors have to be conducting; any nonconducting resistor will disrupt the flow of current (see Fig(\ref{fig:1d})). The value
of $p_c$ for higher dimensional grids is not analytically known. However, when the number of dimensions is high, the answer may be easily approximated.  
For large number of dimensions, a hypercubic lattice in d-dimensions behaves like a Bethe lattice (Fig.(\ref{fig:bethe3})) with $2d$ nearest neighbors, for which the value of $p_c$ is exactly known. For a Bethe lattice with $z$ nearest neighbors, the critical 
value is
\baray
\label{bethe}
p_c = \frac{1}{z-1}
\earay
Since a higher dimensional hypercubic lattice (in $d$-dimensions, for large values of $d$), can be approximated as Bethe lattice
with $2d$ nearest neighbors, the value of the critical probability is 
\baray
\label{pc}
p_c = \frac{1}{2d- 1}
\earay 
Note that as the number of dimensions increases, $p_c$ decreases. \\ 
 
\begin{figure}
\begin{center}
\includegraphics[width=9cm]{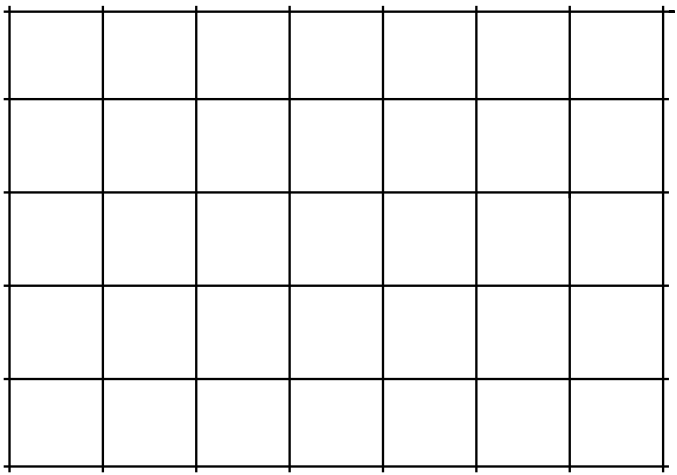}
\vspace{0.1in}
\caption{A square lattice of resistors. All resistors have infinite value.
No current flows.}
\label{fig:nocurrent}
\end{center}
\end{figure}

\begin{figure}
\begin{center}
\includegraphics[width=9cm]{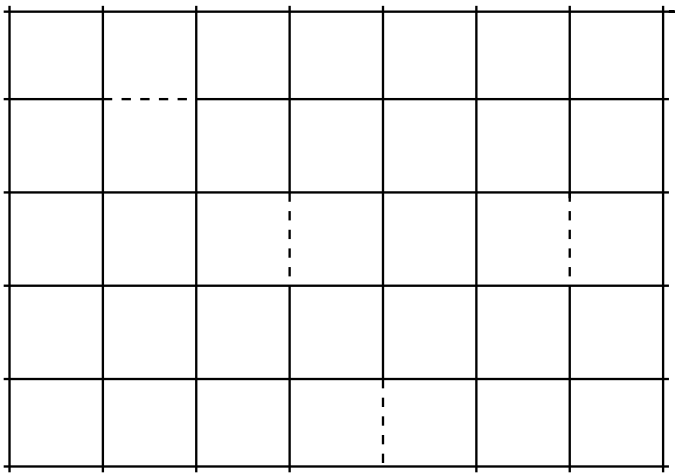}
\vspace{0.1in}
\caption{Solid lines are infinite resistors. Dashed lines are zero resistance
connections. The number of dashed lines are too few to make the grid
conducting, i.e., $p < p_c$.}
\label{fig:subcritical}
\end{center}
\end{figure}

To apply the above idea to the string landscape and efficient tunneling, let us consider a hypercubic lattice of string vacua 
motivated by the Bousso-Polchinski flux landscape \cite{Bousso:2000xa} \footnote{Although our picture is motivated by the Bousso-Polchinski landscape, the mechanism we describe is much more general without any reference to the particular realization of the landscape. In the Bethe lattice, seen in Fig.(\ref{fig:bethe3}), the lattice points would represent the vacua and the lines connecting them would represent the decay channels.} 
This approach lets us talk more generally 
without committing to any particular realization of the landscape. All that is required is individual vaccum sites connected by decay channels -
a situation easily represented by a Bethe lattice. 
We shall be solely interested in efficient tunneling 
and shall consider the usual tunneling
that takes an exponentially large time as ``no tunneling".  As we are interested in the efficacy of efficient tunneling, such an approximation
is valid. Let $p$ be the probability of the occurrence of efficient tunneling. Typically we expect $p$ to be a small number. For resonance
tunneling $p$ will be exponentially suppressed (as discussed in Sec.(\ref{QM})). For DBI tunneling $p$ can be much higher as it naturally
gives a resonance condition for supercritical barriers (Secs.(\ref{efficiency} and \ref{moon})).

The question we
ask is : for a given value of $p$, what is the dimensionality $d$ of the hypercubic lattice (i.e., the number of flux directions, or
in a more general set up simply the number of tunneling channels) for which percolation happens. Naively, as we saw in our discussion of
Bethe lattice, the answer is $1/(2d - 1)$. Hence, for a typical value of $d$ in the $100$ to $1000$ range (corresponding to a hundred to a
thousand flux directions / decay channels), the critical probability
is of the order $\sim 0.01 - 0.001$. This might make DBI tunneling percolate if $p$ is in this range and if the DBI tunneling is
highly enhanced. On the other hand, resonance tunneling has an exponentially suppressed value of $p$ (Sec.(\ref{QM})).     
However, as we will now argue the answer can be exponentially better for resonance tunneling, enough to make it percolate. \\ 

\begin{figure}
\begin{center}
\includegraphics[width=9cm]{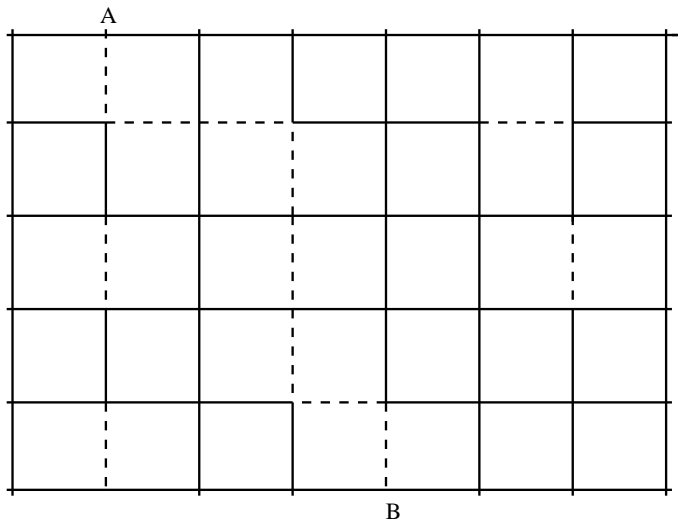}
\vspace{0.1in}
\caption{At critical probability $p_c$, a continuous curve of zero resistance
appears through the lattice. The lattice becomes conducting. Note the 
contnuous dashed line (zero resistance path) from lattice site A to B.
In $2$-dimensions, $p_c = 1/2$.}
\label{fig:critical}
\end{center}
\end{figure}

There is one important difference between the electrical conductivity of the hypercubic lattice of resistors and quantum tunneling in the landscape.
In the grid of resistors, a current can only flow from a given lattice point to a neighboring lattice points. However, in the landscape 
quantum tunneling can occur between non-nearest-neighboring sites as well. The number of sites to which a wavepacket can potentially 
quantum tunnel to can be estimated. Let $\Phi_q$ be the length in the field space over which quantum coherence can be maintained (i.e.
the length of the field space over which the phases of the wavefunctional have definite relationships).  
For example, during inflation, for light fields it is natural to assume that $\Phi_q \sim H$ as they have a natural spread over a Hubble scale.
Let $\Phi_0$ be the typical spacing between neighboring sites in the landscape. Then the number of sites to which a wavepacket can
potentially resonance tunnel to, starting from some initial site, can be approximated as $N \sim (\Phi_q / \Phi_0)^{d}$, where $d$ is the number of directions
about the particular landscape site. This can be an exponentially large number. For the choice of $\Phi_q = 3 \Phi_0$ (i.e. coherence
is maintained over three lattice sites) and $d = 100$, we get $N = 3^{100}$. Now, while using Eq.(\ref{bethe}) to find the critical
probability for percolation to happen, one should use $z \sim N$.

\begin{figure}
\begin{center}
\includegraphics[width=9cm]{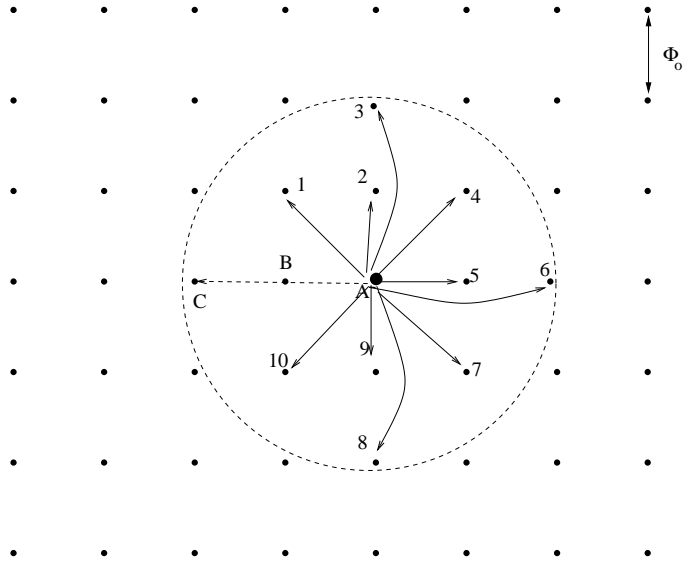}
\vspace{0.1in}
\caption{The dots are the vacua. Starting from vacuum A, quantum tunneling
can potentially happen to all vacua enclosed in the dotted {\it coherence sphere}. Solid directed lines imply that nonresonant tunneling can happen
along the arrow. The dashed directed line implies that
resonance tunneling can happen from A to C via B. $\Phi_o$ is the lattice
spacing.}
\label{fig:coherence}
\end{center}
\end{figure}

\begin{figure}
\begin{center}
\includegraphics[width=5cm]{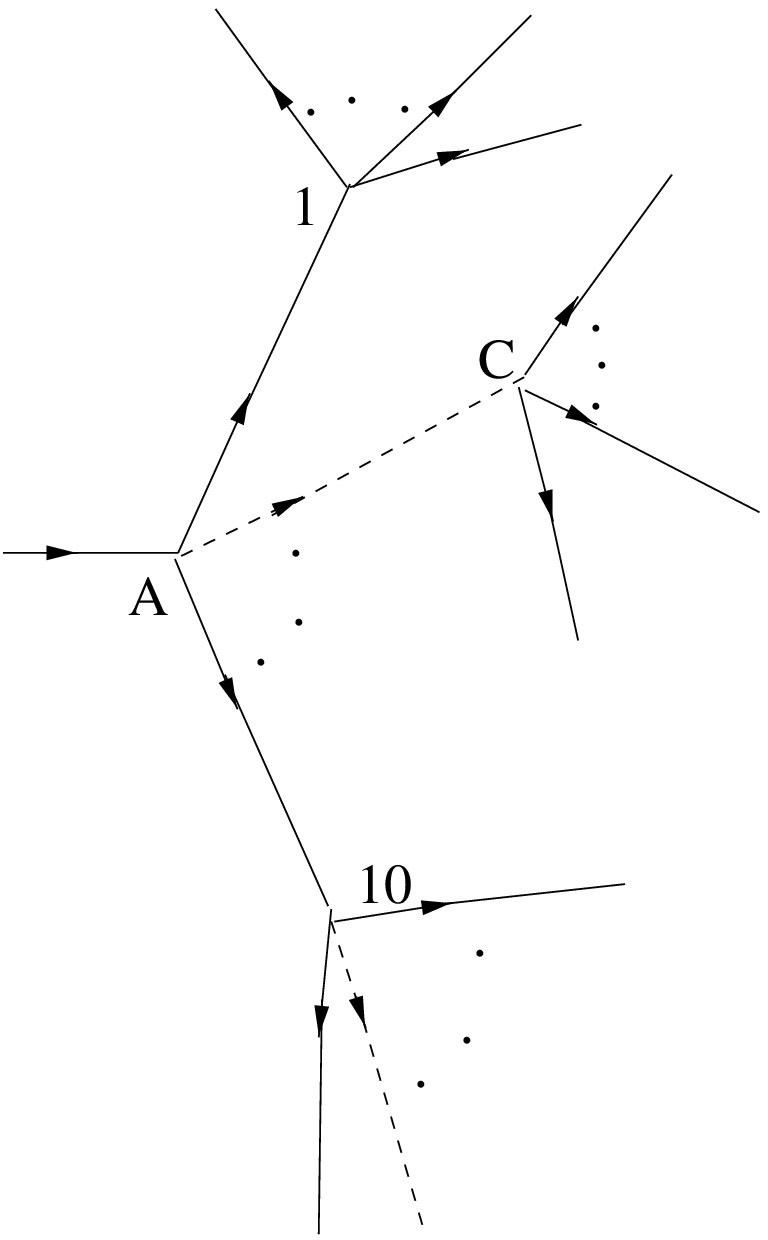}
\vspace{0.1in}
\caption{To apply percolation theory, tunneling from vacuum A to the other vacua inside the coherence sphere in the hypercubic lattice of the previous figure 
 is better represented using a Bethe lattice.  The number of other lattice points connected to a given lattice site is $z$, which is approximately the number of vacua inside the coherence sphere.} 
\label{fig:tree}
\end{center}
\end{figure}

Then we have
\baray
\label{p_c}
p_c \sim \frac{1}{N} \sim \left( \frac{\Phi_q}{\Phi_0} \right)^{-d}
\earay
For the coherence length $\Phi_q$ spanning just a few landscape lattice spacings $\Phi_0$, a very small value of $p_c$ can be achieved
provided $d$ is large.\\

 The scenario discussed in \cite{Tye:2006tg} assumes an exponentially small value of $p$, the probability of the occurrence of efficient
 tunneling. The probability of hitting resonance, $p$, is shown to be of the order of the usual (nonresonance) tunneling probability
 (as discussed in Sec.(\ref{QM})). That is, $p$ is exponentially suppressed. For example, a typical of tunneling probability (using the thin wall Coleman-De Luccia result) in the Bousso-Polchinski scenario is $\sim e^{-100}$ \cite{Bousso:2000xa}. Assuming such a small value of $p$, we see from Eq.(\ref{p_c}) that percolation
can still happen as the critical probabilty $p_c$ can itself be low. There can be a continuous path connecting string landscape sites across which resonance tunneling can happen. The existence of such continuous paths allowing high transition rates between vacua can have important 
cosmological consequence.  Most interestingly, there can be no eternal inflation from such vacua due to
the large number of decay channels, and the decay rate not being so highly suppressed. 

\subsection{Rapid tunneling via the Chebyshev metric, and $\ell^p$ toy landscapes} \label{cheby}
There is another means by which the large dimensionality of the landscape, in combination with efficient tunneling, can yield qualitatively different behavior.  
When a large number of decay channels are present, the total decay rate is written
\be
\Gamma \sim \sum_{q} N_q \exp\left(-27\pi^2 S_{1q}^4/2\epsilon^3\right)\,,
\ee 
where $N_q$ is the number of channels of a given rate (or tension, if we fix $\epsilon$).
For definiteness, we will assume that the tension $S_{1q}$ is of the same order as the vacuum energy difference $\epsilon$.  Each decay channel is therefore very slow, of order $\exp(-27 \pi^2/2) \approx 10^{-58}$.  Qualitatively new behavior occurs when the number of 
accessible final states is large enough to cancel this factor.
Let us assume that the field configuration space has $300$ dimensions, and is a regular square lattice.
We will imagine that the domain wall tension is determined through some measure of distance in the field configuration space, i.e. the single domain walls associated with neighboring vacua are of lower tension than the composite domain walls which bridge to more distant vacua.
Pictorially, $N_q$ is the surface area of a shell in the field configuration space which surrounds the initial vacuum.  

By reformulating the domain wall tension as a function of distance in field space, the physics
is contained in the choice of metric on the (discrete) space of vacua.  We will consider several
different metrics, all being discrete versions of $\ell^p$ norms.  These metrics represent distinct
toy landscapes, where the binding energy of various domain walls (which determines the tension of the composite domain wall) is of the simple form of an $\ell^p$ norm.  The three cases we consider will
be the unbound (or marginally bound) case $p = 1$, the BPS case $p = 2$, and $p = \infty$, known as the``Chebyshev" case.

If one assumes that all of the composite domain walls are marginally bound, i.e. have no binding energy, then 
\be
S_1\left[q_1,q_2,...q_{300}\right] = S_1\times\left(|q_1| +|q_2| +...+|q_{300}|\right)\, ,
\ee
and the shell will be a sphere defined by the ``taxicab metric" on the field configuration space.  If on the other hand the domain walls bind together ({\em \`{a} la} BPS) such that
\be \label{eq:bpswall}
S_1\left[q_1,q_2,...q_{300}\right] = S_1\times\sqrt{q_1^2 + q_2^2 +...+q_{300}^2}\, ,
\ee
then the shells will be spheres defined using the standard metric of ${\bf R}^{300}$.
We will think of the BPS formula for the domain wall tension Eq.(\ref{eq:bpswall}) as a realization of moderately efficient tunneling.  

The more interesting ``Chebyshev" example would have a tension
\be
S_1\left[q_1,q_2,...q_{300}\right] = S_1\times \max(|q_1|, |q_2|, ...,|q_{300}|)\, ,
\ee
in which case the shells are hyper-cubes.  What makes this version of efficient tunneling
interesting is that the volume remains large even when the number of dimensions is large.
In a field configuration space of such large dimension as the landscape (say $300$), there is
a dramatic enhancement for the highly efficient Chebyshev case.  This is because the surface area of a
``taxicab" sphere of radius $p$ is quite small in spaces of large dimensionality.
\be
Vol(S^{299}){\Big \vert}_{taxi} = 2 (2 q)^{299}/300!
\ee
The standard sphere enjoys the larger surface area
\be
Vol(S^{299}){\Big \vert}_{Euclid} = 300 \pi^{150} q^{299}/150!
\ee
but a truly large volume sphere is the Chebyshev sphere:
\be
Vol(S^{299}){\Big \vert}_{Chebyshev} =  300\times 2^{300} q^{299}
\ee
A simple two-dimensional potential which exhibits Chebyshev domain walls is given by
\be
V(\phi_1,\phi_2) = \sin^2(\phi_1) + \sin^2(\phi_2) - 2 \sin^2(\phi_1)\sin^2(\phi_2)\,\,.
\ee

With such a metric $N_q \approx (2q)^{300}$, and it is easy to overcome the large suppression in the Coleman - De Luccia tunneling rate.  If we assume that the individual domain wall satisfies $S_1^4/\epsilon^3 = 1$ then the
total decay rate is given by
\baray
\Gamma &\sim& \sum_{q = 1}^{few} N_q \exp\left(-27\pi^2 q^4/2\right)\\
&\approx& 10^{32}
\earay

This rate is so large that the system would easily percolate.
(Here we use the term `percolate' to mean the nucleated bubbles rapidly coalesce, rather than in the sense of percolation theory used previously.)  Thus instead of eternal inflation in such regions of the landscape, the outcome would more likely resemble a thermalized domain wall foam.  While this toy model is highly speculative, it shows how the large dimensionality of the landscape can conspire with
efficient tunneling to transform eternal inflation into percolation.

\section{Discussion} \label{discuss}
In this paper we have discussed two novel tunneling mechanisms in the string landscape - resonance tunneling
and DBI tunneling. Both lead to an enhancement of the tunneling rate provided certain conditions are met.
We also saw the interesting connection between the two: DBI tunneling for a supercritical barrier can
naturally become resonant. 
Resonance tunneling makes tunneling to far way vacua in the landscape relatively unsuppressed, i.e. far away vacua becomes
accessible and tunneling
becomes efficient. Further, the existence of a large number of decay channels in the landscape can convert this
efficient tunneling into a rapid one. Efficient tunneling is, therefore, a prerequisite for rapid tunneling.
Such rapid modes of tunneling, where the tunneling rates are not exponentially slow compared to the other
time scales in the problem, can have nontrivial
consequences for the evolution of various regions of the universe populating the vacua in the landscape.
As discussed in Sec.(\ref{percolate}) using basic ideas from percolation theory, even if the occurrence of efficient 
tunneling is rare in the landscape,
the existence of a large number of decay channels can make efficient tunneling rapid.  

In the Bousso-Polchinski
scenario, the number of non-collapsing 
decay channels quickly decreases as one approaches the surface of the zero cosmological
sphere (the Bousso-Polchinski sphere) in the lattice of flux vacua.  As pointed out in \cite{Tye:2006tg, Henry},
This reduction in the number of decay
channels for vacua with very low vacuum energy (one of which corresponds to our current vacuum with the low value
of the dark energy) can explain why inspite of efficient tunneling, the universe is stable today.  Paramount to the success of this mechanism is a superselection effect which forbids transition to negative cosmological constant vacuua. 
 The reduction in the
number of decay channels makes efficient tunneling an improbable occurrence. In other words, away from the Bousso-Polchinski
sphere the large number of decay channels can make efficient tunneling natural, whereas close to the Bousso-Polchinski
sphere efficient tunneling can get shut off due to a drop in the number of decay channels. If this transition 
is sudden, then this picture gives a natural explanation for 
a small
cosmological constant. It is important that the shutting-off of rapid tunneling close to the Bousso-Polchinski sphere
be a sudden one, otherwise the intermediate region (where tunneling is slow, i.e. where metastability exists) can
support eternal inflation. \cite{Henry} uses RG group arguments - drawing a parallel with the well studied metal-insulator
transition in condensed matter - to show that such a shut-off of rapid tunneling (which is analogous to the metal-insulator
phase transition) is indeed very sharp. The argument rests on the properties of the beta function and the RG group flow
in the vicinity of the region where the rapid tunneling occurs. One could give similar arguments using percolation theory 
drawing on the connections between percolation and renormalization group (see \cite{Stauffer}).  A superselection sector
boundary at zero cosmological constant has been hypothesized by Banks\cite{Banks:2002nm}.

If efficient and rapid tunneling is a generic feature of the landscape due to a large number of decay channels (away
from the Bousso-Polchinski sphere), then it can interfere with the occurence of eternal inflation. Given 
any vacuum with a large number of decay channels, rapid tunneling will make it decay on a time
scale that is exponentially smaller than that required for the occurence of eternal inflation. 
As efficient and rapid tunnelings constrain eternal inflation, they also constrain the general utility 
of anthropic reasoning. Anthropic reasoning assumes that the vacua in the landscape are generically populated
due to eternal inflation. However, rapid tunneling makes eternal inflation impossible and can make a large
number of landscape vacua unpopulated. 
This could dramatically alter the anthropic reasoning that is often adopted in landscape studies.

As we saw in Sec.(\ref{dbi}), DBI tunneling makes tunneling enhanced when the barriers are of stringy heights.
This is due to the Schwinger effect that becomes operative due to the square root kinetic term (similar
effect is seen of charged relativistic particles in strong electromagnetic field where the action has
a square root kinetic term). After taking account of this relativistic effect, the tunneling rate is 
the same as the Coleman-De Luccia tunneling but with a new potential $V_0^{CDL}(\phi)$ which is related to the
original potential $V_0(\phi)$ by the relation $V_0^{CDL}= V_0(1- fV_0/2)$, $f(\phi)$ being the warp factor.
What is also interesting is that for supercritical barriers ($fV_0 > 2$), DBI
tunneling naturally leads to a resonance condition. Such a connection was certainly not obvious at the outset.
Clearly, this connection between DBI tunneling and resonance tunneling deserves further investigation. In Sec.(\ref{KG}),
we start with the Klein-Gordon wave equation for a relativistic point particle (a zero-dimensional membrane) with 
a certain external potential $V$. We then show that the problem can be mapped to a Schrodinger equation with a 
modified potential $V(1 - V/2)$. This is an alternate way of justifying the simple recipe  $V_0^{CDL}= V_0(1- fV_0/2)$. 

Even though the details of resonance tunneling in quantum field theory are complicated compared to quantum mechanics
because of the infinite dimensional configuration space, in effect all that happens is the constructive interference
in the intermediate vacuum/vacua. In order to meaningfully talk about definite phase relationships over distances
in field space which are involved in the constructive interference, coherence has to be maintained in the field 
disturbances over that field distance. In Sec.~(\ref{percolate}) we have represented this coherence
length as $\Phi_q$ and there is a corresponding notion of the {\it coherent sphere}. During inflation light scalar 
fields have coherence over a field distance $\Phi_q \sim H$. This can
allow for the possibility of constructive interference over distances in the field space that are $\leq H$.

 An important issue in resonance tunneling is what happens to the conventional picture of vacuum decay via bubble
nucleation. In \cite{Coleman:1977py}, the vacuum decay is a first order phase transition that occurs via the
nucleation of various bubbles of various sizes. The bubbles are of subcritical, critical and supercritical sizes.
Subcritical sizes are too small and due to the excess surface energy in the bubble wall they collapse. Critical
bubbles have an exact cancellation of the surface energy and the drop in the volume energy due to the bubble
nucleation. Supercritical bubbles have a large enough size that it is favorable for the bubble to grow. This picture
is bound to change for the decay of the false vacuum via resonance tunneling \cite{Tye:2006tg}.
Resonance tunneling involves the quantum state of the field in the intermediate vacuum - the state corresponding
to the constructive interference. \cite{Tye:2006tg} has argued that subcritical bubbles would nucleate with field
values corresponding to the intermediate vacuum and these small bubbles would collide with each other leaving 
behind the final vacuum and radiation. The slow roll inflation phase could be immediately preceded
by such a phase transition that released radiation. Possible observational signature of such
a thermal phase preceding the slow roll inflation has been investigated in \cite{Sarangi:2006yy}.

In a recent paper \cite{Copeland:2007qf}, the mechanism by which resonance tunneling might proceed in
quantum field theory was investigated. The authors reported a no-go theorem that precludes the occurence of
resonance tunneling as a first order phase transition. The authors , however, have interpreted their result
as a general no-go theorem for resonance tunneling in quantum field theory. We have addressed this no-go theorem
in Appendix \ref{cps}. The complete understanding of what should replace the usual bubble nucleation for
resonance tunneling is lacking. However, as we explain in the appendix, the usual framework that involves an
O($4$) symmetric bounce solution will breakdown.
  
During resonance tunneling, the state of the field in the intermediate vacuum is quantum mechanical. The motion
of the field in this region, in the WKB approximation, is given by a wavefunctional such as Eq.(\ref{allow'}), an
oscillatory solution in the field space. If the intermediate vacuum resembles a quadratic well, then these classical 
solutions correspond to the ground or the excited states of the field in the simple harmonic potential. 
Solutions that are very excited correspond to classical like oscillatory solutions of the field in this potential. 
The field starts at the initial false vacuum, undergoes classical like oscillations in the
intermediate vacuum, and then exits at the final vacuum. It is possible that the oscillatory motion in the 
intermediate vacuum can lead to parametric resonance \cite{Greene:1997fu}. This will lead to particle production.
The backreaction of the particles produced can destroy the resonance condition. It would be interesting
to quantify the conditions under which the backreaction is large enough to interfere with the resonance tunneling.  

Efficient and rapid tunneling can have interesting phenomenological consequences \cite{rapid} \footnote{See also \cite{Mersini} 
and references therein.}.
It will be interesting to do a detailed analysis such as \cite{Clifton:2007bn} with rapid tunneling taken into account.
One such scenario involving domain walls has been studied in \cite{Davoudiasl:2006ax}. Typically, since
tunneling is assumed to be an exponentially slow process, any signs of nearby vacua are relegated to
the far past when the universe was eternally inflating. This makes any possibilities of finding observational
signatures of the string landscape impossible even in principle. 
However, efficient tunneling that becomes rapid can change this picture and provide some handle on nearby vacua.
In particular, while the inflaton field sources inflation of the universe, there could be other light
fields that have typical de Sitter spreads in their vacuum expectation values (a fluctuation of the order
of a Hubble mass scale during one Hubble time). This spread in the
vevs can lead these light fields to explore near by vacua (the coherent sphere discussed in Sec.(\ref{percolate}) is
related to this spread in the vev of the light fields) and display rapid tunneling if the conditions are conducive. 
Assuming that rapid
tunneling happens during the last $60$ e-folds of inflation, different Hubble patches of the postinflationary
universe could end up in different vacua. These vacua will be separated by domain walls whose tensions depend
on the structure of the local landscape that is sampled by the efficient tunneling. Depending on the distribution
of the heights of the barriers, there could be a distribution in the tensions of the domain walls resulting from
efficient tunneling. As discussed in \cite{Davoudiasl:2006ax}, requiring the domain walls to not pose any
cosmological problems can lead to constraints on the structure of the local landscape (of course, this is
under the assumption that efficient tunneling happens during the last $60$ efolds). The vacua on either side of
the domain wall will generically differ in their vacuum energies. If this split in vacuum energies is large enough,
the domain walls will experience a pressure gradient and will attain relativistic speeds very quickly. This
can help get rid of domain walls. Requiring the domain walls to go away before big bang nucleosynthesis (BBN)
requires the split in vacuum energies to be greater than MeV$^4$ and a lower bound on the domain wall energy scale 
of the order $10$ TeV$^4$.  However, if the split in the vacuum energies is less than MeV$^4$, the walls
will stay around and eventually collapse into primordial black holes. The mass of the black hole will depend on
the domain wall tension. A spread in the domain wall tensions will lead to a spread in the masses of the
primordial black holes. Domain walls with higher tensions will lead to black holes with lower masses which, in turn, evaporate first. When the black hole evaporates it converts the energy which was initially in the
domain wall into radiation. This leads to a reheating phase of the universe. In \cite{Davoudiasl:2006ax} it
is shown that, starting with domain walls of various tensions that are formed right after inflation, 
as soon a domain wall of a particular tension becomes the dominant form of energy in the
universe, it collapses to black holes. These black holes eventually evaporate away. The lighter black 
holes will evaporate first.
This can lead to multiple phases of reheating in the early universe. Requiring the reheat temperature 
to be high enough for successful BBN requires the wall tensions to be above $10^{-5} M_P$.  Since
the domain walls (and the black holes to which they collapse) become the dominant form of energy
in the universe at the time of the collapse, it is not possible for these primordial black holes
to be dark matter candidates.

Finally, it is plausible that both modes of rapid tunneling might serve as alternate solutions to the empty universe problem, which
one encounters in trying to explain the small cosmological constant using relaxation mechanisms \cite{Bousso:2000xa,
Brown:1987dd}. The final stage of the sequence of sampling of different sites of the landscape
should correspond to the very low cosmological constant that we observe 
today. However, tunneling typically takes an exponentially large time 
and, when implemented to solve the cosmological constant problem (a la
Brown-Tietelboim, Bousso-Polchinski) has the associated problem of 
the  Empty Universe. The original Brown-Tietelboim mechanism had two problems
: the gap problem and the empty universe problem. The Bousso-Polchinski 
proposal succeeded in solving the gap problem due to the large number of directions that exist in the flux landscape, 
and gives plausible scenarios for solving the empty universe problem. The exponentially large time required for tunneling 
from one vacuum site to another dilutes
away the matter and radiation components of the universe and leads to an empty
universe - this is the empty
universe problem. However, as we have discussed in this paper, tunneling need not always be an exponentially slow process.
While tunneling usually takes an exponentially large time to take place, when certain conditions
tunneling rates can be highly enhanced. Further, the large number of directions present in the landscape lead to a 
large number of decay channels, some of which might lead to rapid tunneling (as discussed in Sec.(\ref{percolate})).
In scenarios where such efficient and rapid tunneling might be generic, tunneling
rate can be unsuppressed and the cold empty universe may be avoided. 

A relevant issue while discussing preferred paths (percolation) in the landscape is that of
the initial condition \footnote{We thank Puneet Batra for detailed discussions regarding the importance 
of the initial conditions.}.  While in our analysis of resonance tunneling in Secs.(\ref{QM}, \ref{QFT})
we solve a time independent Schr\"odinger equation, the problem of tunneling is a dynamical one.
One should prepare a wave packet of the particle/field and solve the corresponding time dependent 
Schr\"odinger equation. This is a standard exercise in quantum mechanics, one starts with an ansatz 
$\psi(x,t) = \phi(x) e^{iEt}$ for the wave packet and reduces the time dependent equation to a time
independent one. The results obtained from the time independent analysis are then recovered. 
Perfect resonance condition is satisfied only in certain $\it bands$
$\Delta E$ of the energy of the wave packet. The wavepacket itself has a spread in energy $\delta E$. For
the complete passage of the entire wave packet through the resonance band the requirement is $\delta E < \Delta E$.
This will ensure that even without a large number of decay channels, the efficient tunneling is rapid. 
What decides the width of the wave packet in the landscape? Presumably this question is tied up
with the question of initial conditions which has been much debated \cite{Vilenkin:1998rp}. Simple applications of quantum
cosmology lead to very different results about the initial condition of the universe \cite{init}.
As discussed in \cite{Firouzjahi:2004mx} this might be a problem of staying
within the minisuperspace approximation. Going beyond the minisuperspace scheme can solve some pathologies
of the wavefunction of the universe. Luckily, in our picture we do not need to know the width of the wavepacket.
Resonance ensures that the tunneling is efficient. And the presence of a large number of decay channels can
make the efficient tunneling rapid.

In our analysis of resonance tunneling in QFT, we have not included gravity. In our semiclassical
analysis we have assumed a fixed flat spacetime background. Admittedly gravity should be 
included while discussing tunneling in the string landscape. Inclusion of gravity
will include a gravitational Hamiltonian along with the usual matter Hamiltonian ($H = H_m + H_g$) 
in the Schr\"odinger equation. This Schr\"odinger equation is simply the Wheeler-De Witt equation
($H \Psi = 0$) for the wavefunctional of the gravitational and the matter degrees of freedom expressing
the diffeomorphism invariance. Now $\Psi$ is the wavefunction of the universe. However, we do not expect
gravity to change the qualitative picture. For vacuum sites with a large vacuum energy, the effect
of gravity is not important \cite{Coleman:1980aw}. Gravity becomes important for vacuum sites with small
vacuum energy by making them stable.

\acknowledgments

We thank Puneet Batra, Brian Greene, Dan Kabat, Louis Leblond, Sarah Shandera, Henry Tye
and Alexander Westphal
for invaluable discussions. We would also like to thank Edmund Copeland, 
Antonio Padilla and Paul Saffin for vigorous discussions regarding
resonance tunneling in QFT.     
SS would like to thank Jiajun Xu for key 
discussions at the beginning of the work. 
BS would like to thank Shanta de Alwis and Oliver DeWolfe for helpful conversations.  SS and GS thank 
Hooman Davoudiasl for numerous stimulating discussions.
BS and SS thank Adam Brown and Amanda Weltman for helpful conversations.
This work was supported in part by the DOE
under contracts DE-FG02-92ER40699 (SS),
DE-FG02-95ER40896 (GS) and DE-FG02-91ER40672 (BS),
NSF CAREER Award PHY-0348093 (GS), a Research Innovation
Award (GS) and a Cottrell Scholar Award (GS) from Research Corporation.

\appendix

\section{Appendix: The thin wall approximation and the landscape} \label{appx1}

\begin{figure}
\begin{center}
\includegraphics[width=9cm]{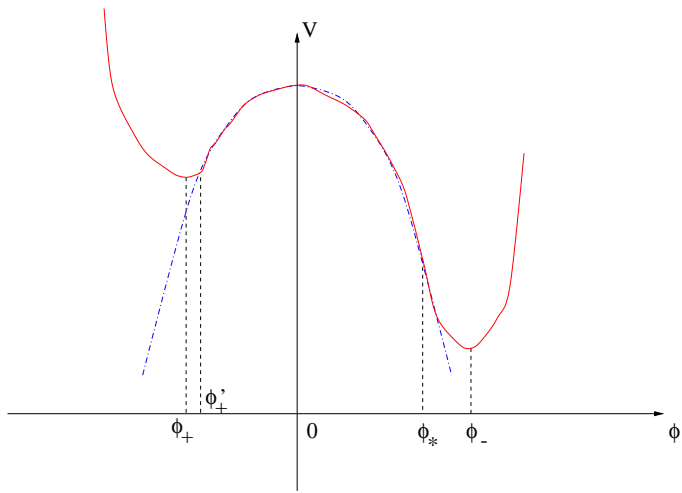}
\vspace{0.1in}
\caption{The red (continuous) line shows the potential $V(\phi)$ with a metastable vacuum at $\phi_+$ and
a true vacuum at $\phi_-$. The blue (dashed) line is a hypothetical inverted quadratic potential $V = V_0 
- \frac{1}{2}\mu^2\phi^2$ chosen so as to fit the actual potential $V(\phi)$ as closely as possible 
in the field range between $\phi_+$ and $\phi_-$.}
\label{quadratic}
\end{center}
\end{figure}

We have used the thin wall approximation throughout this paper and one might be concerned
regarding the thin wall approximation when the vacua and their energy
differences span large intervals \footnote{We thank Alexander Westphal for detailed
discussions about the robustness of the thin wall approximation.}. The energy difference between
two neighboring vacua, $\epsilon$, will in general not be small in the landscape. However,
the thin wall approximation can still be used when $\epsilon$ is not exceedingly small.
In this appendix we show that the thin wall formula has a wide range of applicability
by deriving a rough estimate of the upper bound on $\epsilon$ (in units of
the height of the barrier $V_{barrier}$) that still allows a thin wall description. In
particular, we shall show that one can stay within the thin wall approximation and
still have $\epsilon \sim V_{barrier}$. We are interested in an order of magnitude 
estimate. Consider a potential $V(\phi)$ with metastable minimum at $\phi_{+}$ and a 
true minimum at $\phi_{-}$ (see the red curve Fig.(\ref{quadratic})). The energy difference between
the two vacua is $\epsilon = V(\phi_+) - V(\phi_-)$. We assume that the potential $V(\phi)$ has a shape
that allows a thin wall approximation for small $\epsilon$. We investigate what happens as $\epsilon$ 
is increased starting with a small value well within the thin wall regime. 
 
\begin{figure}
\begin{center}
\includegraphics[width=7cm]{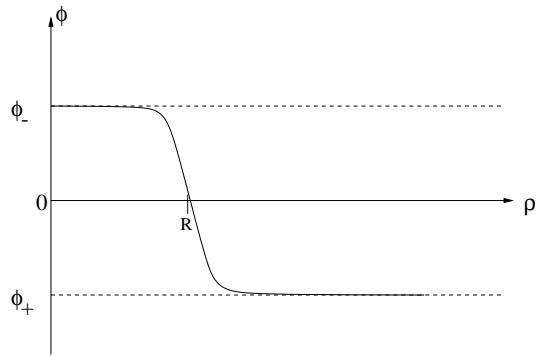}
\vspace{0.1in}
\caption{A thin wall bounce configuration. The radius $R$ of the bubble wall is much greater than
the thickness of the wall (the range of $\rho$ over which the configuration changes over from 
$\phi_-$ to $\phi_+$).}
\label{thinwall}
\end{center}
\end{figure}

 For small $\epsilon$, the thin wall picture of bubble
nucleation holds, i.e. the wall of the bubble is thin (compared to the radius of the bubble) and 
at the moment of the nucleation of the bubble, the field configuration outside the bubble is 
$\phi = \phi_+$ and inside the bubble it is $\phi = \phi_-$. This configuration (the bounce) is obtained by
solving the Euclidean equation of motion (using $\rho$ for the radial Euclidean coordinate)
\ba
\label{bounce'}
\frac{d^2\phi}{d\rho^2} + \frac{3}{\rho}\frac{d\phi}{d\rho} - V'(\phi) = 0
\ea
which describes the motion of the field in an inverted potential $-V(\phi)$. The boundary conditions
for the bounce solution are : $\frac{d\phi}{d\rho}(\rho = 0) = 0$, and $\phi(\infty) = \phi_+$.
In the thin wall approximation the friction term, $\frac{3}{\rho} \frac{d\phi}{d\rho}$, is neglected.
The bounce solution $\phi(\rho)$ begins near the true vacuum $\phi = \phi_-$ at $\rho = 0$, stays
very close to the true vacuum till it reaches the bubble wall at the radius $\rho=R$, then it quickly
crosses the wall (the time over which it does so is given by the inverse curvature of the potential at its top) and goes very close to the false vacuum $\phi = \phi_+$ and reaches this value 
at $\rho = \infty$. A configuration such as this is shown in Fig.(\ref{thinwall}).  Over the wall
region the field quickly rolls over the valley of the inverted potential $-V(\phi)$ (i.e. the top of the
potential $V(\phi)$). The thickness of the wall is given by $\delta \sim \mu^{-1}$ where $\mu^2 =
|V_{top}''(\phi)|$ (i.e., $\mu$ is the curvature of the top of the potential barrier). The thin wall
condition is therefore : $R = 3S_1/\epsilon >> \mu^{-1}$. \\
 
\begin{figure}
\begin{center}
\includegraphics[width=7cm]{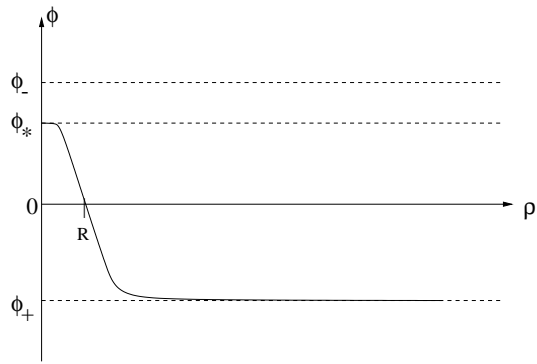}
\vspace{0.1in}
\caption{A thick wall bounce configuration. At $\rho = \infty$, $\phi = \phi_+$. As $\rho$
decreases, the field value tends towards $\phi_-$ but never quite attains that value. At
$\rho = 0$ (center of the bubble), $\phi = \phi_*$. The value of $\phi_*$ is obtained
by solving the equation of motion for the bounce  with appropriate
boundary conditions, and satisfies the condition $\phi_+ < \phi_* < \phi_+$ for our
potential. The radius $R$ of the bubble wall is the same as
the thickness of the wall.} 
\label{thickwall}
\end{center}
\end{figure}

 Now we start increasing $\epsilon$. There will be a gradual transition from the thin wall regime
to the thick wall regime. For large values of $\epsilon$ the bounce will take the thick wall
configuration described in Fig.(\ref{thickwall}). At $\rho = \infty$, the field will sit
at its false vacuum value $\phi_+$. As we start moving towards the center at $\rho = 0$, the
field will tend towards the true vacuum $\phi_+$ but will never attain that value. Instead
for all $\epsilon$ above a certain value, the field configuration at $\rho = 0$ will be 
$\phi(0) = \phi_*$ where $\phi_*$ can be uniquely determined by solving the Euclidean equation
of motion (Eq.(\ref{bounce'}) with the boundary conditions $\phi(\rho=\infty)= \phi_+$, 
and $\frac{d\phi}{d\rho}(\rho = 0) = 0$ \footnote{Intuitively it is clear that once $\epsilon$ is
very large, the decay rate will not depend on $\epsilon$ any more. There is no way for the field
sitting at the false vacuum $\phi_+$ to gauge the depth of the true vacuum on the other side of
the barrier prior to tunneling. In other words, once $\epsilon$ exceeds some critical value which is
large, the decay rate will simply become insensitive to the exact location of the true vacuum.}.
At the nucleation of the (now thick walled) bubble, the field at the center of the bubble will
be at $\phi = \phi_* $. Post nucleation, as the bubble grows, $\phi$ will classically roll from
$\phi_*$ towards $\phi_+$.

   Next we find a rough estimate for $\phi_*$, and the corresponding value of energy 
difference $\epsilon_* = V(\phi_+) - V(\phi_*)$. The thin wall approximation breaks 
down much before 
$V(\phi_+)-V(\phi_-) = \epsilon_*$. We shall determine this value $\epsilon_{tw}$ shortly.
The reader should note the difference between $\epsilon_{tw}$ (above which thin wall is
a bad approximation), and $\epsilon_*$ (above which the decay rate becomes very insensitive
to $V_{-}$. To find $\epsilon_*$
we adopt the following strategy. We fit the potential $V(\phi)$  with an 
inverted quadratic potential $V = V_0 - \frac{1}{2}\mu^2\phi^2$ (blue line in 
Fig.(\ref{quadratic}))
\footnote{This analysis is therefore applicable to those potentials that can be fitted with an inverted quadratic potential in the region between the metastable and the true vacua. One could do a similar analysis for the case where the potential is fitted with an inverted potential of polynomial
$n$ (i.e., cubic for $n = 3$, etc). One shall find (and we have checked this numerically) that the thin wall approximation holds for larger and larger $\epsilon$ as $n$ is increased.}. For such a model potential the 
Euclidean equation of motion is linear and exactly solvable. Let $\phi'_+$ be the point on 
this inverted quadratic potential that is the closest to the true vacuum $\phi_+$ of the actual 
potential $V(\phi)$ that we are interested in. We assume that our original potential $V(\phi)$ has
a shape that allows the existence of such a $\phi'_+$ very close to $\phi_+$. Now we solve the 
Euclidean equation of motion (Eq.(\ref{bounce'})) using this inverted quadratic potential 
\ba
\label{bounce''}
\frac{d^2\phi}{dz^2} + \frac{3}{z}\frac{d\phi}{dz} + \mu^2 \phi = 0
\ea
where $z$ is a Euclidean coordinate. If we interpret $z$ as a time coordinate, then this equation 
describes the motion of the field in the inverted potential $-V_0 + \frac{1}{2}\mu^2\phi^2$.
Starting from an intial value $\phi(z=0) = \phi_{initial}$, the field will roll down the potential, 
pass through the minimum at $\phi = 0$, and come to rest at some field value $\phi_{final}$ (and 
then roll back towards the origin, but we are not interested in
this motion of the field after it comes to rest at $\phi_{final}$). $\phi_*$ is defined as that
initial value $\phi_{initial}$ for which $\phi_{final} = \phi'_+$.
The solution to Eq.(\ref{bounce''}) (with the boundary condition $\phi(z=0)=\phi_*$) is
\ba
\label{soln}
\phi(z) = 2\phi_* \frac{J_1(\mu z)}{\mu z}
\ea
where $J_{n}$ is the Bessel function of the first kind of order $n$. The field begins with zero 
speed at $\phi_*$. When its speed vanishes again at the other side of the potential (at $z = z'_+$, 
say), its field value is $\phi'_+$. It can be shown (using the recursion relations $2\frac{dJ_1(z)}{dz}
= J_0(z)-J_2(z)$, and $2J_1(z)=z(J_0(z)+J_2(z))$) that this happens when $J_2(z=z'_+) = 0$.
This gives the values $z \simeq 5.14$, $\phi'_+ \simeq \phi_* J_0(5.14) \simeq -0.13 \phi_*$. 
Since we assume $\phi_+$ and $\phi'_+$ are very close to each other, we get $\phi_*$ in terms
of $\phi_+$ as 
\ba
\label{phi_*}
\phi_* \simeq -7.7 \phi_+
\ea
The corresponding vacuum energy difference is
\ba
\label{e_*}
\epsilon_* = \frac{1}{2}\mu^2|\phi_+^2 - \phi_*^2| \simeq 58.3 V_{barrier}
\ea
where $V_{barrier} = V(0) - V(\phi_+)$ is the height of the barrier infront of the 
false vacuum $\phi_+$. For $\epsilon > \epsilon_*$, the bounce becomes 
insensitive to the true vacuum at
$\phi_-$. For $\epsilon > \epsilon_*$, on bubble nucleation, the field value at 
the center of the bubble is $\phi_*$. It is interesting that the value of $\epsilon_*$ 
can be of the order of ten times the height of the barrier. 

 The thin wall approximation has, however, broken down much earlier. To work with in
thin wall approximation, the wall thickenss ($\mu^{-1}$) much be much less than the
radius of the bubble ($R = 3S_1/\epsilon$), i.e.
\ba
\epsilon < \epsilon_{tw} = 3S_1\mu
\ea
For the inverted quadratic potential we can calculate $\epsilon_{tw}$. 
$S_1$ is given by
\ba
S_1 = \int_{\phi_1}^{\phi_+}d\phi \sqrt{V(\phi) - V(\phi_+)} \nonumber \\
= \frac{V_0}{\mu}\left( \sin^{-1}\left(\frac{\mu\phi}{\sqrt{2V_0}}\right) + \frac{\mu^2\phi^2}{2V_0} \sqrt{\frac{2V_0}
{\mu^2\phi^2}-1}\right)_{\phi_-}^{\phi_+}
\ea
For $\phi_+ \simeq \phi_- \simeq \pm \sqrt{2V_0}/ \mu $, we get $S_1 \simeq \pi V_0/\mu$,
i.e., $\epsilon_{tw} = 3S_1\mu \simeq 3\pi V_0$. So $\epsilon$ can be of the order of $V_{barrier}$
(in this particular case, since the two vacua are close to $V(\phi) = 0$, $V_0 \sim V_{barrier}$ ).

\section{Appendix: Evading the CPS no-go theorem} \label{cps}

After the initial appearance of this paper, Copeland, Padilla
and Saffin produced a theorem in ref.\cite{Copeland:2007qf} that
analyzes the possibility of resonance tunneling in quantum field
theory. Starting with a set of assumptions, the authors show that resonance 
tunneling cannot be a first order phase transition process 
in a single scalar field quantum field theory. This follows from the authors'
argument that the ``bound state'' condition (which is a prerequisite
for resonance tunneling) cannot happen in field theory.  There are
various ways to circumvent this no-go theorem (as mentioned in
\cite{Copeland:2007qf} itself). In this section
we explain their proof and show that while the proof is valid if the
all the assumptions hold, the assumptions are inapplicable. 
The very framework that the authors employ to
prove the no-go theorem breaks down in the context of resonance tunneling. 

We first briefly explain the CPS no-go theorem.
As explained in Sec. \ref{resonance} , the occurence of resonance 
tunneling requires that the resonance condition be satisfied, i.e.
\baray
\label{reson2:2} 
W = \left( n + \frac{1}{2} \right) \pi \,\,,
\earay
where $n = 0,1,2,3,...$. $W$ is related to  
the characteristics of the intermediate vacuum
\baray
\label{W22}  
W = \int_{\lambda_2}^{\lambda_3} d\lambda' \sqrt{2m(\lambda')[E - U_{eff}(\lambda')]} \,\,.
\earay
$\lambda_2$ and $\lambda_3$ being the turning points (see Sec \ref{resonance} 
for derivation). Eq.(\ref{reson2:2}) is precisely the condition for 
realizing a bound state of the scalar field in the intermediate vaccum.
Once this condition is satisfied, the transmission probability given by
\baray
\label{ItoV:2}
T_{I \to V} = 4 \left( \left( \Theta \Phi + \frac{1}{\Theta \Phi}\right)^2 \cos ^2 W + \left( \frac{\Theta}{\Phi} + \frac{\Phi}{\Theta} \right)^2  \sin^2 W \right)^{-1}\,\,.
\earay
is enhanced. If the two barriers are shaped such that $\Theta = \Phi$, then the transmission probability is order unity.

It is possible to construct an O($4$) symmetric bounce when discussing tunneling across a single barrier. One works entirely in the
Euclidean regime (the under-the-barrier region) and finds the bounce solution. Analytically continuing it to the Lorentzian region
gives the O($3,1$) growing bubble solution. Eq.(\ref{ItoV:2}), however, describes the transition from one vacuum to another which lacks
an entirely Euclidean description. The intermediate vacuum is in the Lorentzian region and contributes the oscillatory contributions
of $\cos^W$ and $\sin^2 W$. The Euclidean O($4$) bounce solution will not be present anymore due to these Lorentzian interventions
from the intermediate vacuum. The usual description of the phase transition via O($4$) symmetric bounce is no more valid.

Now we explain the CPS no-go theorem. Consider a $1+1$ dimensional
field theory for simplicity (the arguments are easily generalizable to more dimensions). 
The theorem depends on the following  assumptions
\begin{enumerate}
\item the bound state condition applies, i.e. $W=(n+1/2)\pi$.
\item it satisfies and classical equations of motion and is not the false vacuum.
\item it obeys the boundary conditions : $\phi(r=0) = \phi_{true}$, $\phi(r=\infty) = \phi_{false}$.
\item it has zero energy.
\item $d\phi/dt$ vanishes at two different times $t_1$ and $t_2$.
\end{enumerate}
Starting with these assumptions, the authors construct an integral $I(x)$ given by
\ba
I(x) = \int_{t_1}^{t_2}dt \left( V(\phi) - \frac{1}{2}\left( \frac{\partial \phi}{\partial t}\right)^2  - \frac{1}{2}\left( \frac{\partial \phi}{\partial x}\right)^2 \right)
\ea
Because of assumptions $2$ and $5$, $dI/dx = 0$. So $I$(at any point)$= I(\infty) = 0$ (since the field configuration at $\infty$ is assumed
to be the false vacuum, $I(\infty)$ vanishes). The energy functional is given by
\ba
E = \int d^3x \left( V(\phi) + \frac{1}{2}\left( \frac{\partial \phi}{\partial t}\right)^2  + \frac{1}{2}\left( \frac{\partial \phi}{\partial x}\right)^2 \right),
\ea
and by assumption $4$ $E=0$. Combining the two results $I=0$ and $E=0$ then implies the following
\ba
\int dt d^3x \left(\frac{1}{2}\left( \frac{\partial \phi}{\partial t}\right)^2  + \frac{1}{2}\left( \frac{\partial \phi}{\partial x}\right)^2 \right)
= \int dt d^3x V(\phi) = 0.
\ea
That is, subject to the above assumptions, if the field configuration is sitting at the false vacuum at infinity, it is doomed to sit at the false vacuum every where else. 

The above no-go theorem uses several well known facts about instantons in quantum field theory.
Because the imaginary part of the vacuum energy density (and thus the decay rate per unit
volume) is proportional to $\exp(-S_E)$, the Euclidean action of the instanton must be finite.  This requires
the instanton to asymptote to the false vacuum.   Another property of {\em resonance tunneling} instantons
(at least in quantum mechannics) is that they possess a classically allowed region where the field could
oscillate in {\em real} time.  Because classical oscillation implies the existence of two turning points, CPS
point out that in some coordinates, the field configuration describing the expanding bubble (which is also the
instanton) should have two ``spacial" slices  $t_1$ and $t_2$ over which the field is ``at rest."  (We use quotations
because it is difficult to translate the analogous quantum mechanical problem to quantum field theory.)

The mathematical machinery of the CPS theorem then proceeds to make the following argument:
Since the spacial slices must extend to spacial infinity, they must extend into a region where the field is
asymptotic to the false vacuum.  Furthermore, it can be elegantly shown that any field configuration which
obeys the equations of motion in some four-volume $V_4$ is uniquely determined by its value (and derivative) on $\partial V_4$.
Thus since the field is constant on the two spacial slices $t_1$ and $t_2$, as well constant at spacial infinity, the field
must be simply the false vacuum solution of zero Euclidean action, and not the instanton we were attempting to find.
Since the instanton satisfying the stated conditions does not exist, there is no associated tunneling event.

The way we can evade this theorem is to point out that by continuity, the ``classical region in which the field can oscillate" must be contained within the domain wall.  This is simply due to the fact that the ``classical" region of the potential is between the false and true vacua.  Since the domain wall is by definition the region where the field interpolates between the false and true vacua, the ``classical region" must be localized to a thin slice within the domain wall itself.
This calls into question the assumption that the ``spacial slices" $t_1$ and $t_2$ do in fact asymptote to
regions containing false vacuum.  Because $t_1$ and $t_2$ must be localized within the domain wall, they are by
definition far (in field space) from the false vacuum.  

We have not shown the existence of resonance tunneling in quantum field theory beyond stating the 
necessary conditions that are to be satisfied (in Sec. \ref{QFT}). However we have shown the CPS
theorem does not demonstrate the impossibility of this task.



\end{document}